\documentclass[fleqn,usenatbib]{mnras}

 \pdfoutput=1 

\usepackage{newtxtext,newtxmath}

\usepackage[T1]{fontenc}

\DeclareRobustCommand{\VAN}[3]{##2}
\let\VANthebibliography\thebibliography
\def\thebibliography{\DeclareRobustCommand{\VAN}[3]{##3}\VANthebibliography}

\usepackage{ae,aecompl}
\usepackage{xparse,xspace}

\usepackage{cancel}
\usepackage{siunitx}
\usepackage{graphicx}	
\usepackage{amsmath}	
\usepackage{amsfonts}
\usepackage{graphicx}
\usepackage{color}
\usepackage{bm}
\usepackage{lipsum}
\usepackage{nccmath}
\usepackage{tabularx}
\usepackage{textcase}
\usepackage{caption}
\usepackage{multicol}
\usepackage{relsize}
\usepackage{multirow}
\usepackage{graphicx}
\usepackage[export]{adjustbox}
\usepackage{pdflscape}
\usepackage{CJKutf8}
\begin{document}

\title[Biosignatures Below Radius Valley with Twinkle]{Is LTT 1445 Ab a Hycean World or a cold Haber World? Exploring the Potential of \textit{Twinkle} to Unveil Its Nature}


\author[Phillips et al.]{
Caprice Phillips$^{1}$
\thanks{E-mail: phillips.1622@buckeyemail.osu.edu},
Ji Wang\begin{CJK*}{UTF8}{bkai}(王吉)\end{CJK*}$^{1}$, 
Billy Edwards$^{2,3,4}$, 
Romy Rodr\'iguez Mart\'inez$^{1}$, 
\newauthor Anusha Pai Asnodkar$^{1}$,   B. Scott Gaudi$^{1}$\\
$^{1}$Department of Astronomy, The Ohio State University, Columbus, OH, 43210,USA\\
$^{2}$Blue Skies Space Ltd., 69 Wilson Street, London, UK\\
$^{3}$Universit\'{e} Paris-Saclay, Universit\'{e} Paris Cit\'{e}, CEA, CNRS, AIM, 91191, Gif-sur-Yvette, France\\
$^{4}$Department of Physics and Astronomy, University College London, Gower Street, London, UK\\
}

\date{Accepted XXX. Received YYY; in original form ZZZ}

\pubyear{2022}


\label{firstpage}
\pagerange{\pageref{firstpage}--\pageref{lastpage}}
\maketitle

\begin{abstract}
We explore the prospects for \textit{Twinkle} to determine the atmospheric composition of the nearby terrestrial-like planet LTT 1445 Ab, including the possibility of detecting the potential biosignature ammonia (NH$_{3}$). At a distance of 6.9 pc, this system is the second closest known transiting system and will be observed through transmission spectroscopy with the upcoming \textit{Twinkle} mission. \textit{Twinkle} is equipped with a 0.45 m telescope, covers a spectral wavelength range of 0.5 - 4.5 $\mu$m simultaneously with a resolving power between 50 - 70, and is designed to study exoplanets, bright stars, and  solar system objects.  
 We investigate the mission's potential to study LTT 1445 Ab and find that \textit{Twinkle} data can distinguish between a cold Haber World (N$_2$-H$_2$-dominated atmosphere) and a Hycean World with a  H$_2$O-H$_2$-dominated atmosphere, with a $\chi_{\nu}^{2}$ = 3.01. Interior composition analysis favors a Haber World scenario for  LTT 1445~Ab, which suggests that the planet probably lacks a  substantial water layer.
We use \texttt{petitRADTRANS} and a \textit{Twinkle}  simulator to simulate transmission spectra for the more likely scenario of a cold Haber World for which NH$_{3}$ is considered to be a biosignature. We study the detectability under different scenarios: varying hydrogen fraction, concentration of ammonia, and cloud coverage. We find that ammonia can be detected  at a $\sim$ 3$\sigma$ level for optimal (non-cloudy)  conditions with 25 transits and a volume mixing ration of 4.0 ppm of NH$_{3}$. We provide examples of retrieval analysis to constrain potential NH$_{3}$ and H$_{2}$O in the atmosphere. Our study illustrates the potential of \textit{Twinkle} to  characterize atmospheres of potentially habitable exoplanets. 
\end{abstract}


\section{Introduction}
\label{sec:introduction}
\textit{Twinkle} is an upcoming space-based telescope with a 0.45 m primary aperture and a broad visible to infrared wavelength coverage (0.5 -- 4.5 $\mu$m). The \textit{Twinkle} space mission~\citep{TwinkleSPIE} will conduct two simultaneous surveys during its first three years of operation, which is scheduled to begin in 2025. While one of these will focus on studying objects within our own solar system, the other will be dedicated to the study of extrasolar targets. A large portion of the latter survey will be used to study exoplanet atmospheres, the science case for which \textit{Twinkle} was originally conceived~\citep{twinkle_exo}. There are nearly 900 confirmed transiting exoplanets within \textit{Twinkle's} field of view, as well as over 1400 planet candidates from the Transiting Exoplanet Survey Satellite (TESS, \citealt{Ricker2015}), offering the potential for a structured population survey of exoplanet atmospheres.

\textit{Twinkle} can be  highly complementary to the James Webb Space Telescope (JWST). While JWST will deliver unprecedentedly precise data, there will be limited time allocated to exoplanet sciences. Therefore, it is likely to only be used to observe the most exciting targets. To this end, \textit{Twinkle} can provide low-resolution spectroscopy to provide an initial atmospheric characterization to promote further study or be used to refine planetary and orbital parameters. Moreover, certain JWST instruments/modes cannot observe bright targets due to saturation limits, and \textit{Twinkle} can fill in the gap for bright targets. Furthermore, the planets studied with \textit{Twinkle} can be methodically selected, building up large sets of data with specific goals in mind whereas each JWST proposal often focuses only on a small number of worlds. Combining \textit{Twinkle} data with the JWST mission will allow us to achieve a more comprehensive picture of exoplanet atmospheres. 

\par
The Kepler Space Mission \citep{borucki2010} has shown that super-Earths/mini-Neptunes are amongst the most abundant type of planet (\citealt{fressin2013}; \citealt{fulton2017}). There is an observed gap in the distribution of these planet sizes, known as the radius valley (\citealt{fulton2017}; \citealt{VanEylen2018}). Below the radius valley ($<$ 1.5 R$_{\oplus}$), these planets are known as super-Earths/terrestrial-like planets. Studies have investigated their ability to hold onto a hydrogen-based atmosphere due to their decreased mass and decreased surface gravity from both ground-based (\citealt{DiamondLower2018}; \citealt{DiamondLowe2020}) and space-based observatories~\citep{Edwards2021}.

Planets with H$_{2}$/He dominated atmospheres may be more amenable targets for transmission spectroscopy with upcoming space-based missions such as \textit{Twinkle}. The presence of H$_{2}$ can raise the scale height and therefore the transmission signal features for observations~(\citealt{Miller_Ricci_2008}; \citealt{HuOceans2021}).  H$_{2}$-dominated atmospheres may also produce different biosignatures, such as NH$_{3}$ in cold Haber Worlds~\citep{Seager2013a}.

\par 
In this work we assess the detectability of the potential biosignature ammonia on the terrestrial-like planet, LTT 1445 Ab with the upcoming \textit{Twinkle} space mission.  We first provide a summary of previous literature on ammonia as a potential biosignature in \S \ref{sec:ammonia_biosignature}. We then describe the target selection process for the study in \S \ref{sec:selection_criteria}. The process to distinguish LTT 1445~Ab from a cold Haber World or Hycean World is described in \S \ref{sec:HabervsHycean}. Major findings on the detectability of NH$_{3}$ are presented in  \S \ref{sec:mainresults}. Finally, we present our retrieval analysis to support the major findings in \S \ref{sec:retrieval_analysis} and conclude in \S \ref{sec:conclusions}.

\section{Ammonia as  a Potential Biosignature}
\label{sec:ammonia_biosignature}

\cite{Seager2013a} first proposed NH$_3$ as a biosignature gas in a H$_{2}$ and N$_{2}$ dominated atmosphere -- nicknamed a cold Haber World~(e.g. \citealt{Seager2013b};  \citealt{HuangAmmonia2021}). Cold Haber Worlds are named after the Haber-Bosch process which is the main industrial process for producing NH$_{3}$ from N$_{2}$ from the air, H$_{2}$ and an iron catalyst, combined with high temperatures and pressures. The reaction is as follows: $$3H_{2} + N_{2} \longrightarrow 2NH_{3}$$
\par
{Since the proposal of NH$_{3}$ as a biosignature, there have been a multitude of studies to investigate its detectability with JWST and future Extremely Large Telescopes~(e.g. \citealt{Chouqar2020}; \citealt{Wunderlich2020}; \citealt{Phillips2021}; \citealt{Rajan2022}}).
\cite{Phillips2021} explored the detection of the potential biosignature NH$_{3}$ in gas dwarfs, exoplanets with radii between Earth and Neptune with potentially H$_{2}$ dominated atmospheres. They found that a minimum of 0.4 ppm would be needed to detect the ammonia features in transmission spectroscopy with the NIRSpec and NIRISS (SOSS) instruments/modes on JWST, given optimal cloud-free atmospheric conditions. \cite{HuangAmmonia2021} assessed ammonia as a potential biosignature in terrestrial-like planets and found that a minimum of 5.0 ppm ammonia in the atmosphere would be needed to be detectable by JWST using the NIRSpec/G395M mode for the 3.0 $\mu$m ammonia feature. 
\par

{Although NH$_{3}$ is a strong candidate biosignature in H$_{2}$ and N$_{2}$ atmospheres, there is still need to consider the potential of false positives.} An overview of false positives of NH$_{3}$ was provided by \citealt{Seager2013b} and \citealt{Catling2018}, alongside thesis work by Evan Sneed\footnote{\url{https://scholarsphere.psu.edu/resources/6c6f6ce8-3a94-40f5-a895-4165556b0f58}}. Recently, \cite{HuangAmmonia2021} laid out a few examples of minor abiotic sources of NH$_{3}$ for Earth/terrestrial-like planets including: trace components in volcanic gas eruptions, iron doping TiO$_{2}$ containing sands, and lightning.
\section{Target Selection}
\label{sec:selection_criteria}

\begin{figure*} 
     \centering
     \includegraphics[width=9cm]{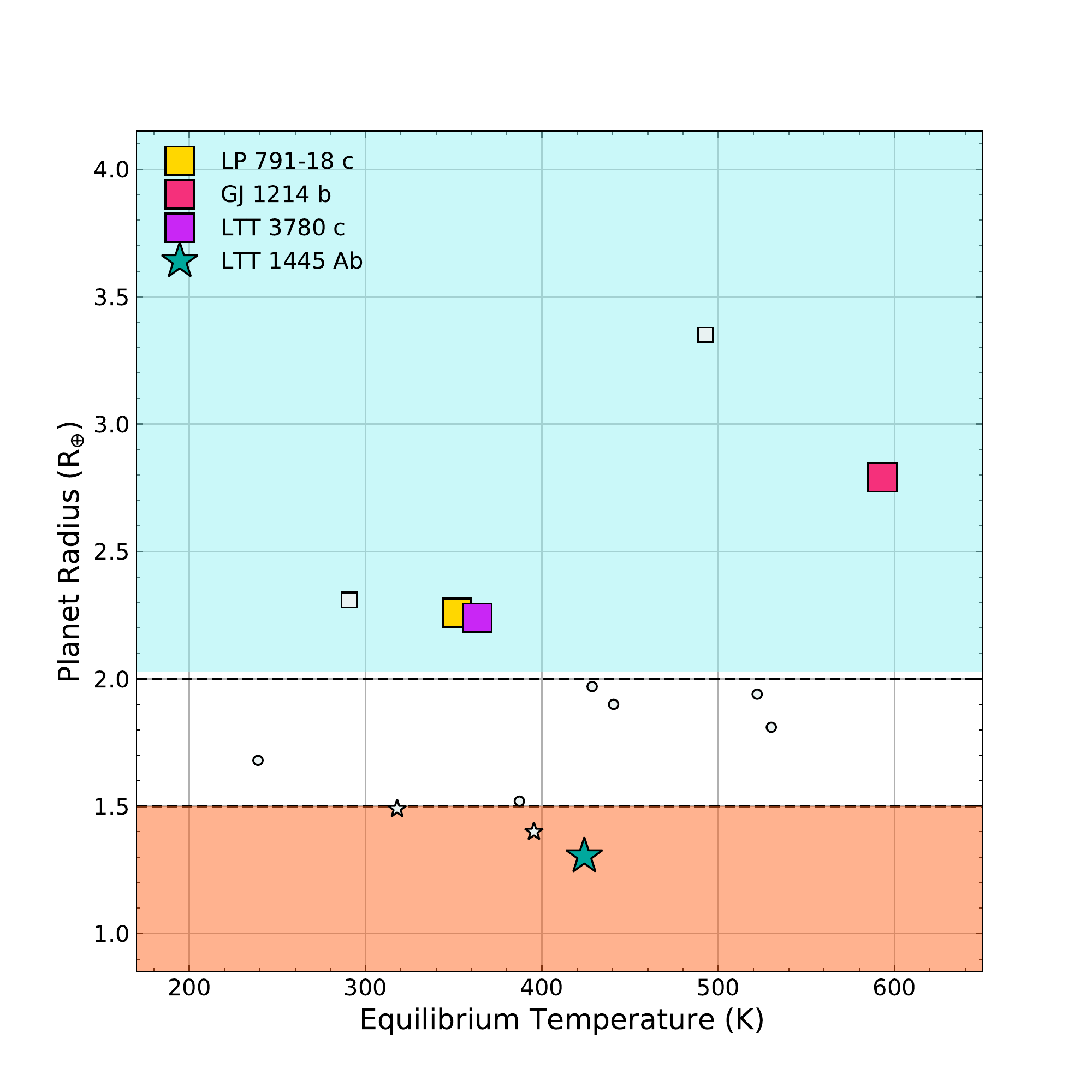}\includegraphics[width=9cm]{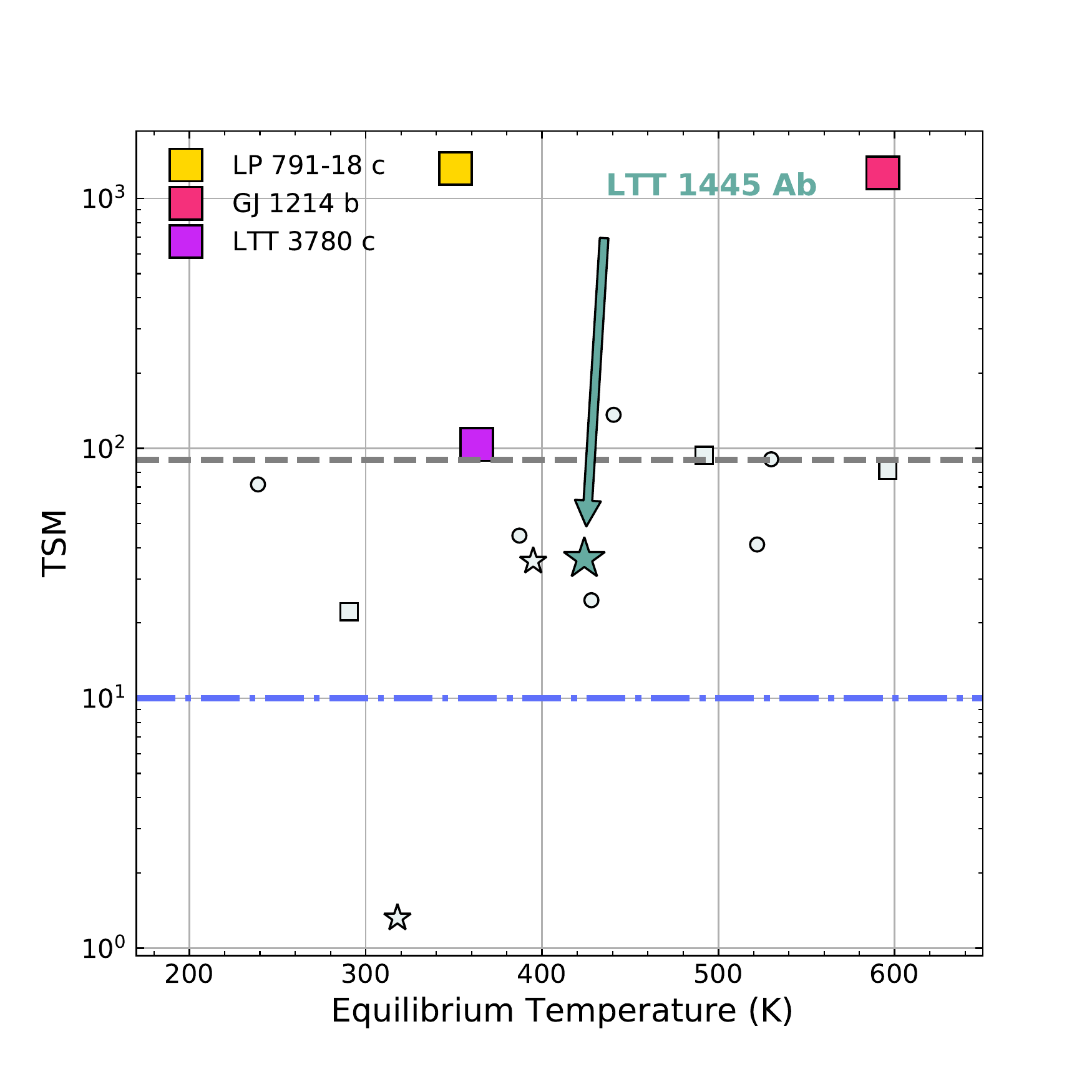}
    \caption{\textbf{Left:} Planet radius (R$_{\oplus})$ vs equilibrium temperature (K) for \textit{Twinkle} objects that meet initial selection criteria (1-3). The orange shaded region and star shapes are objects  below the radius valley (<1.7 R$_{\oplus}$). Objects within the radius valley (1.5 -- 2.0 R$_{\oplus}$) are marked with circle shapes. The blue shaded regions and square markers are targets above the radius valley (>2.0 R$_{\oplus}$). The target of interest for this study LTT 1445 Ab is marked with a blue-green star. \textbf{Right:} The  modified transmission spectroscopy metric (TSM) for H-band defined by \citet{Kempton2018}. There are larger planets above the radius valley with the highest TSM: GJ 1214~b (pink square), LTT 3780~c (purple square), and LP 791-18~c (gold square). However, these targets have low S/N estimates with \textit{Twinkle }for 25 transits. The  horizontal grey line represents the recommended TSM threshold for targets (R = 1.5 -- 10.0 $R_{\oplus}$) to be considered to be selected for high quality atmospheric characterization~\citep{Kempton2018}. The purple dotted-dashed line represents the threshold TSM values (TSM $>$ 10) for terrestrial like planets~($<$1.5 R$_{\oplus}$).}
    \label{fig:selection_criterion}
\end{figure*}

We explore possible targets of interest for characterization that are within the field of view for \textit{Twinkle}. Targets are evaluated using the following criteria: (1) planet radii between 1.3 and 3.4 R$_{\oplus}$, (2) equilibrium temperature (T$_{eq}$) below 650 K, (3) distance within 50 pc, (4) an initial S/N estimation (<S/N> $\geq$ 3$\sigma$) for \textit{Twinkle} using TwinkleRad~\citep{twinkle_orbital_tool} for a baseline of 25 transits and (5) a modified transmission spectroscopy metric (TSM) from \cite{Kempton2018} (see Equation \ref{eq:tsm_adapted}).

Compared to the work in \cite{Phillips2021}, we use an expanded parameter space of radii and equilibrium temperatures. \cite{nixon2021deep} found that the phase structure of water-rich sub-Neptunes show indication that planets with a H/He envelope could host liquid H$_{2}$O in the liquid phase at up to 647 K at pressures of $218$ to $7\times$10$^{4}$ bar. We also explore a slightly lower radius space (1.3 R$_{\oplus}$), as \cite{HuangAmmonia2021} evaluated ammonia as a promising biosignature on terrestrial-like planets (e.g. a 1.75 R$_{\oplus}$, and 10 M$
_{\oplus}$ exoplanet around an M-dwarf). 

We search the NASA Exoplanet Archive\footnote{\url{https://exoplanetarchive.ipac.caltech.edu/}} for targets that meet our criteria. We find that LTT 1445~Ab has the highest (modified) TSM for terrestrial targets below the radius valley~(Figure \ref{fig:selection_criterion}). We also explore targets that lie within the radius valley, but find low S/N estimates for NH$_{3}$ detection given the baseline of 25 transits. Initially objects such as GJ 1214~b, LP 791-18~c, and LTT 3780~c meet the first three criteria for target selection and produce high TSM metrics for planets above the radius valley. However, these targets either have known flat transmission spectra \citep{Kreidberg2014} and/or currently have low S/N estimates for \textit{Twinkle}. We therefore focus on LTT 1445~Ab in subsequent analyses.  
\begin{figure}
    \centering
    \includegraphics[width=\columnwidth]{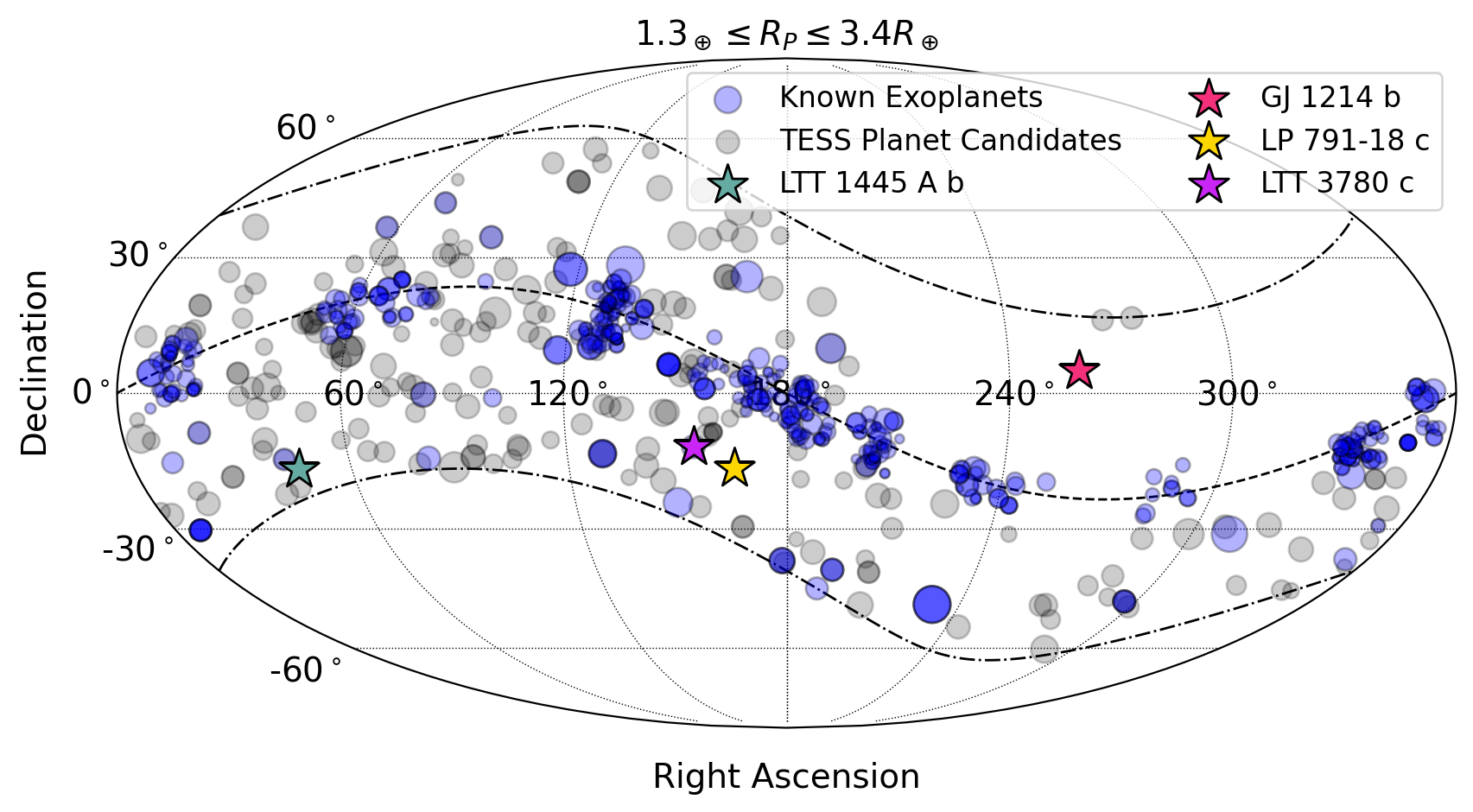}
    \caption{Field of view showing known exoplanets (blue circles) with planetary size (1.3$_{\oplus}$ $\leq$ R$_{p}$ $\leq$ 3.4$_{\oplus}$ ). The $TESS$ planet candidates are shown in the gray markers. The size of the markers correspond to the host star's K-band magnitude. The target of interest for this paper LTT 1445~Ab is shown with a star marker. GJ 1214 b, LP 791-18 c, LTT 3780~c are shown with a pink, gold, and purple star respectively.}
    \label{fig:FOV_Twinle}
\end{figure}
\subsection{LTT 1445 Ab}
Since LTT 1445~Ab is a potential target, we provide a brief introduction of the system. LTT 1445~Ab lies at a distance of 0.038 AU from its host star and has an orbital period of 5.4 days~\citep{Winters2019}. The host system is comprised of three mid-to-late M dwarfs. The host star, LTT 1445~A, is bright (K$_{s}$ = 6.50 mag). LTT 1445~A is also the closest M-dwarf to host a transiting planet~\citep{Winters2021}, making this system a prime target for atmospheric characterization. A summary of key stellar and planetary parameters is shown in Table \ref{tab:planetary_parameters}. During the first three years of operations, \textit{Twinkle} will conduct an extrasolar survey \citep{TwinkleSPIE}. We use the tool from \citet{twinkle_orbital_tool} to determine that, during the time frame of this survey, there will be 29 transits available for observation with \textit{Twinkle}.

\begin{table}
    \centering
    \caption{Planetary and stellar parameters for LTT 1445 Ab from 
   \label{tab:planetary_parameters}
    \citealt{Winters2021}}
    \begin{tabular}{lc} 
    \hline
    \hline
    M$_{p}$ (M$_{\oplus}$)&2.87$^{+0.26}_{-0.25}$\\
    R$_{p}$ (R$_{\oplus}$)&1.304$^{+0.067}_{-0.060}$\\
    T$_{eq}$(K)&424$\pm$21\\
    Distance (pc)&14.98$\pm$0.01\\
    H-band$_{s}$ (mag)&6.774$\pm$0.038\\
    T$_{s}$ (K)&3337$\pm$150\\
    logg (dex)&3.217$^{+0.050}_{-0.053}$\\
    t$_{14}$(hrs)&1.367$^{+0.017}_{-0.016}$\\
    Fe/H (dex)&-0.34$\pm$0.08\\
    Eccentricity&0.19$^{+0.35}_{-0.14}$\\
    \hline\hline
    \label{tab:LTTparams}
    \end{tabular}
\end{table}

Despite the relative small size of LTT 1445 Ab, it is a target of interest for atmospheric studies. \cite{Winters2021} calculate a TSM of 30 for LTT 1445 Ab which is higher than those for LHS 1140~b~\citep{Dittmann2017} and TRAPPIST 1-f~\citep{Gillon2017}. We implement a modified version of the Kempton TSM (Equation \ref{eq:tsm_adapted}), 

\begin{equation}
\label{eq:tsm_adapted}
\rm{TSM} = (\rm{Scale \  factor}) \times \frac{R_{p}^{3}T_{eq}}{M_{p}R_{\star}^{2}} \times 10^{-m_{x}/5}
\end{equation}
In Equation \ref{eq:tsm_adapted}, $T_{eq}$ is the planet equilibrium temperature in Kelvin, $R_{p}$ is the planet radius is Earth radii, $M_{p}$ is the planet mass in Earth mass, $R_{\star}$ is the host star radius in solar radii, and $m_{x}$ is the apparent magnitude of the host star.

The Scale factor in Equation \ref{eq:tsm_adapted} is designed to be a normalization constant to give near-realistic S/N values for 10 hr observing with the JWST/NIRISS instrument \citep{Kempton2018}. The scale factor is different for planets with R < 1.5 R$_{\oplus}$ (scale factor = 0.190) and planets with  1.5 R$_{\oplus}$ < R < 2.75 R$_{\oplus}$ (scale factor = 1.26). For more details about the method used and scale factor determination, see \citet{Kempton2018}. 

We adopt the TSM equation for the \textit{Twinkle} wavelength coverage of 0.4 -- 4.5 $\mu$m coverage, for the H-band coverage and L-band coverage. The H-band region is approximately the  central wavelength for Channel 0 for \textit{Twinkle} which covers 0.5 - 2.4 microns. The L-band region is approximately the  central wavelength for Channel 1 for \textit{Twinkle} which covers 2.4 - 4.5 microns. We find a H-band TSM of 36.0~(Figure \ref{fig:selection_criterion}) and L-band TSM of 44.0.

\section{A Haber World vs.  A Hycean World}
\label{sec:HabervsHycean}
In this section we explore the scenario of LTT 1445 Ab as a Hycean world~\citep{MadhusudhanHyean2021} and its implications for detection with \textit{Twinkle}. A Hycean world is defined as a planet that has a water-rich interior with massive oceans underneath a H$_{2}$-dominated atmosphere~\citep{MadhusudhanHyean2021} We also use the bulk properties (M$_{p}$ and R$_{p}$) of LTT 1445~Ab to constrain the interior composition and to assist with distinguishing a cold Haber World from a Hycean World. 

As shown in \S \ref{sec:twinkle-distinguish}, our interior composition analysis indicates that LTT 1445~Ab is likely not a Hycean world. As a result, we model LTT 1445~Ab as a cold Haber World.

\subsection{Simulating a cold Haber World Spectrum}
\par
 We use the Python package, \texttt{petitRADTRANS}\footnote{\url{https://gitlab.com/mauricemolli/petitRADTRANS}} \citep{molliere2019}, to simulate the planetary atmosphere for transmission spectroscopy. Similar to \citet{Phillips2021}, we utilize the low resolution mode and use a 1-D Gaussian Kernel to smooth the spectra to the  \textit{Twinkle} Channel 0 (0.5 -- 2.4 $\mu$m) and Channel 1 (2.4 -- 4.5 $\mu$m) resolution of R = 70 and R = 50 respectively.  The reference pressure is set to $P_{0}$ = 1.0 bar. We consider lower pressures in the atmosphere in this work, expanding from the lower bound of the P-T profile of Earth.  We use lower pressure value of 10$^{-9}$ bar, opposed to a lower 10$^{-6}$ bar pressure from an adjusted P-T profile of Earth~\citep{Phillips2021}.
\par

We modeled the performance of \textit{Twinkle} using TwinkleRad, an adapted version of the radiometric tool described in \citet{mugnai_arielrad}. We note that, due to the ongoing detailed design work, there are currently significant performance margins built-in to this simulator.
\par
\begin{figure}
    \centering
     \includegraphics[width=1.0\columnwidth]{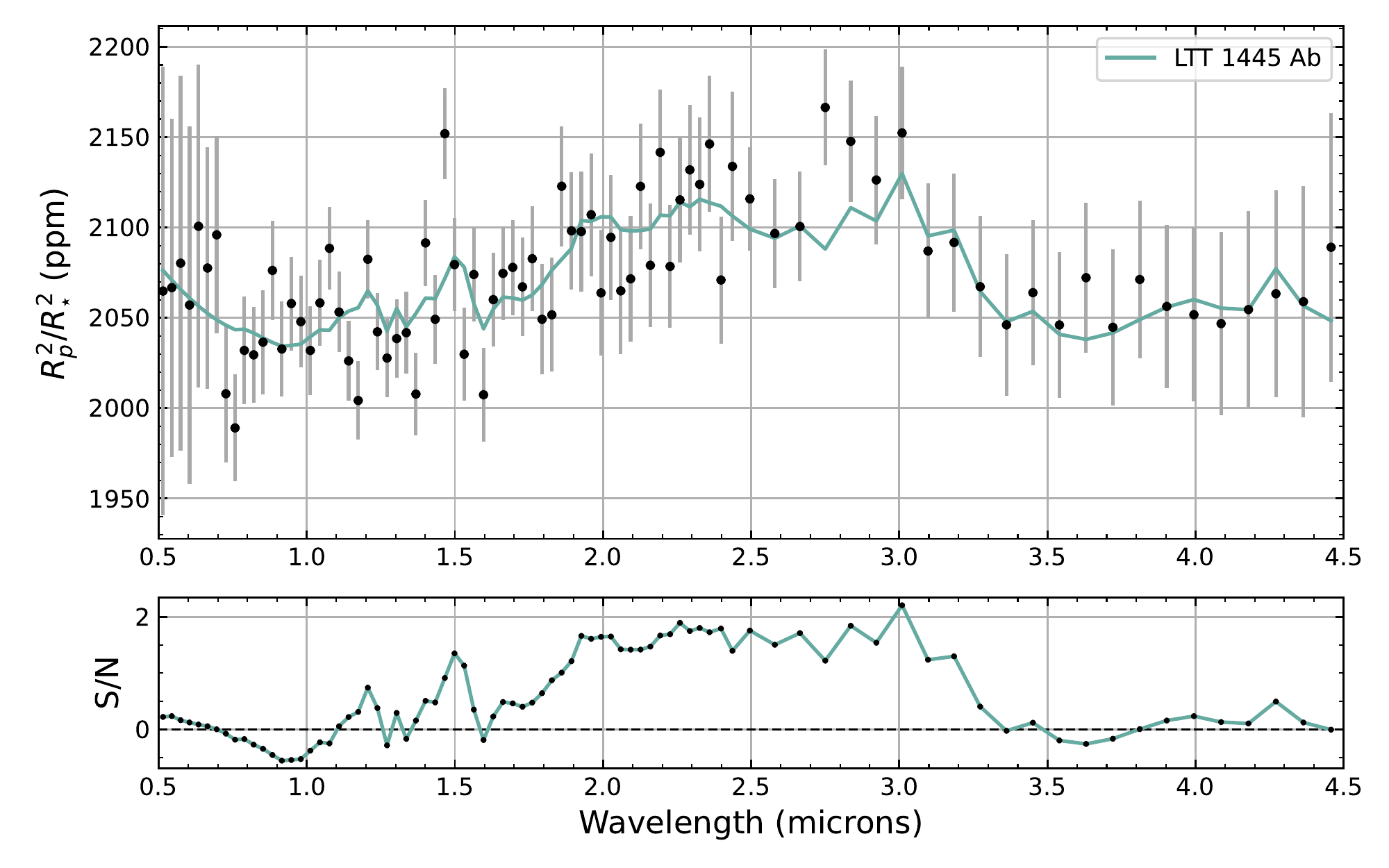}
    \caption{\textbf{Top:} Simulated cold Haber World transmission spectra of LTT 1445 Ab with 25 transits. Low mean molecular weight case ($\mu$ $\sim$ 4.5).  \textbf{Bottom:} The corresponding S/N }
    \label{fig:transmission_Twinkle}
\end{figure}

 We employ the same methods used in \cite{Phillips2021} to calculate the atmospheric composition for a cold Haber world. We use the the values in Table \ref{tab:superearth_low} to build the synthetic spectrum with a base atmosphere of 90 per cent H$_{2}$ and 10 per cent N$_{2}$. A fixed number of 25 transits is set to determine NH$_{3}$ is detectable. \texttt{TauREx 3}~\citep{AlTaurex2021} is used to rebin the spectra. Synthetic noise is added using a random Gaussian distribution.  The S/N detection metric and threshold (<S/N> $\geq$ 3$\sigma$) is the same as in \citet{Phillips2021}. The simulation for 25 transits with \textit{Twinkle} with the corresponding S/N is shown in Figure \ref{fig:transmission_Twinkle}.
 
\begin{table}
\centering
\caption{Species used for \texttt{petitRADTRANS} to generate synthetic spectra with 90 per cent H$_{2}$  and 10 per cent N$_{2}$ atmosphere for a cold Haber World. \textbf{Notes:} (a) We adopt the mixing ratios for gases other than N$_{2}$ or H$_{2}$~\citep{Seager2013b}, assuming no major change in the chemistry  (b) N$_{2}$ has no rotational-vibrational transitions, so there are no signatures observable at infrared wavelengths, so this feature is not available in \texttt{petitRADTRANS} but is used to determine the mean molecular weight of the atmosphere}
\label{tab:superearth_low}.
\begin{tabular}{ccc}
\hline\hline
Species & VMR$^{a}$ & MMR\\
\hline
H$_{2}$O &9.17$\times$10$^{-7}$&3.62$\times$10$^{-6}$\\
CO$_{2}$&2.90$\times$10$^{-9}$&2.81$\times$10$^{-8}$\\
CH$_{4}$&2.90$\times$10$^{-8}$&1.02$\times$10$^{-7}$\\
H$_{2}$&8.25$\times$10$^{-1}$&3.62$\times$10$^{-1}$\\
CO&9.17$\times$10$^{-10}$&5.64$\times$10$^{-9}$\\
OH&9.17$\times$10$^{-16}$&3.42$\times$10$^{-15}$\\
HCN&9.17$\times$10$^{-10}$&5.44$\times$10$^{-9}$\\
NH$_{3}$&3.6$\times$10$^{-6}$&1.37$\times$10$^{-5}$\\
N$_{2}$$^{b}$&9.17$\times$10$^{-2}$&5.64$\times$10$^{-1}$\\
He&8.25$\times$10$^{-2}$&7.25$\times$10$^{-2}$\\
\hline
\end{tabular}
\end{table}

\subsection{Can Twinkle Distinguish a cold Haber World from a Hycean World?}
\label{sec:twinkle-distinguish}

While LTT 1445 Ab is not in the traditional habitable zone of its star~\citep{Winters2019}, it is considered as a candidate Hycean world where a liquid ocean may exist underneath a H$_{2}$-dominated atmosphere.  In a Hycean world scenario, LTT 1445~Ab lies in the Hycean habitable zone. The Hycean habitable zone is defined as regions corresponding to  the maximum irradiation that allows for habitable conditions at the surface of the ocean~\citep{MadhusudhanHyean2021}. For a Hycean world,  \cite{MadhusudhanHyean2021} consider H$_{2}$O, CH$_{4}$, and NH$_{3}$ as  potentially abundant molecules in  a H$_{2}$-based atmosphere. 
The question remains if we can use \textit{Twinkle} to distinguish between a Hycean world and a cold Haber world. This is an important question because detecting NH$_{3}$ needs to be put into larger context of which world it belongs to. For a cold Haber world, NH$_{3}$ is regarded as a biosignature~(\citealt{Seager2013a}), however, NH$_{3}$ detection may be unrelated to life in a Hycean world but nonetheless is a critical chemical species in the atmosphere~\citep{MadhusudhanHyean2021}.  
 
In order to distinguish between two worlds, we first simulate \textit{Twinkle} observations of a Hycean world for LTT 1445~Ab. Then, we compare the simulated data with that of a cold Haber world using a reduced $\chi^{2}$ statistic. Below we detail the two steps. 

We simulate the Hycean-case scenario for LTT 1445 Ab by adopting the volume mixing ratios provided in \cite{MadhusudhanHyean2021}. In their work, they assume a volume mixing ratio of 0.1, 5.0 $\times$10$^{-4}$, and 1.3 $\times$ 10$^{-4}$ for H$_{2}$O, CH$_{4}$, and NH$_{3}$ respectively. We then convert these values to mass mixing ratios (Table \ref{tab:hycean}). We use these mass mixing ratios as input to \texttt{petitRADTRANS} and simulate the atmosphere. The pressure bar is set to P$_{0}$ = 1.0 bar, low resolution mode is used, and an  424 K isothermal atmosphere is assumed.

\begin{table}
    \centering
    \caption{Species used for \texttt{petitRADTRANS} to generate synthetic Hycean-world spectra atmosphere, adopted from \citet{MadhusudhanHyean2021}}
    \label{tab:hycean}
    \begin{tabular}{lcc} 
    \hline\hline
    Species & VMR & MMR \\
    \hline
    H$_{2}$O &7.26$\times$10$^{-2}$&2.39$\times$10$^{-1}$\\
    CH$_{4}$&2.90$\times$10$^{-8}$&1.02$\times$10$^{-7}$\\
    H$_{2}$&6.54$\times$10$^{-1}$&2.39$\times$10$^{-1}$\\
    NH$_{3}$&3.66$\times$10$^{-6}$&1.37$\times$10$^{-5}$\\
    \hline\hline
    \end{tabular}
\end{table}

\par
\begin{figure*}
    \centering
    \includegraphics[width=\textwidth]{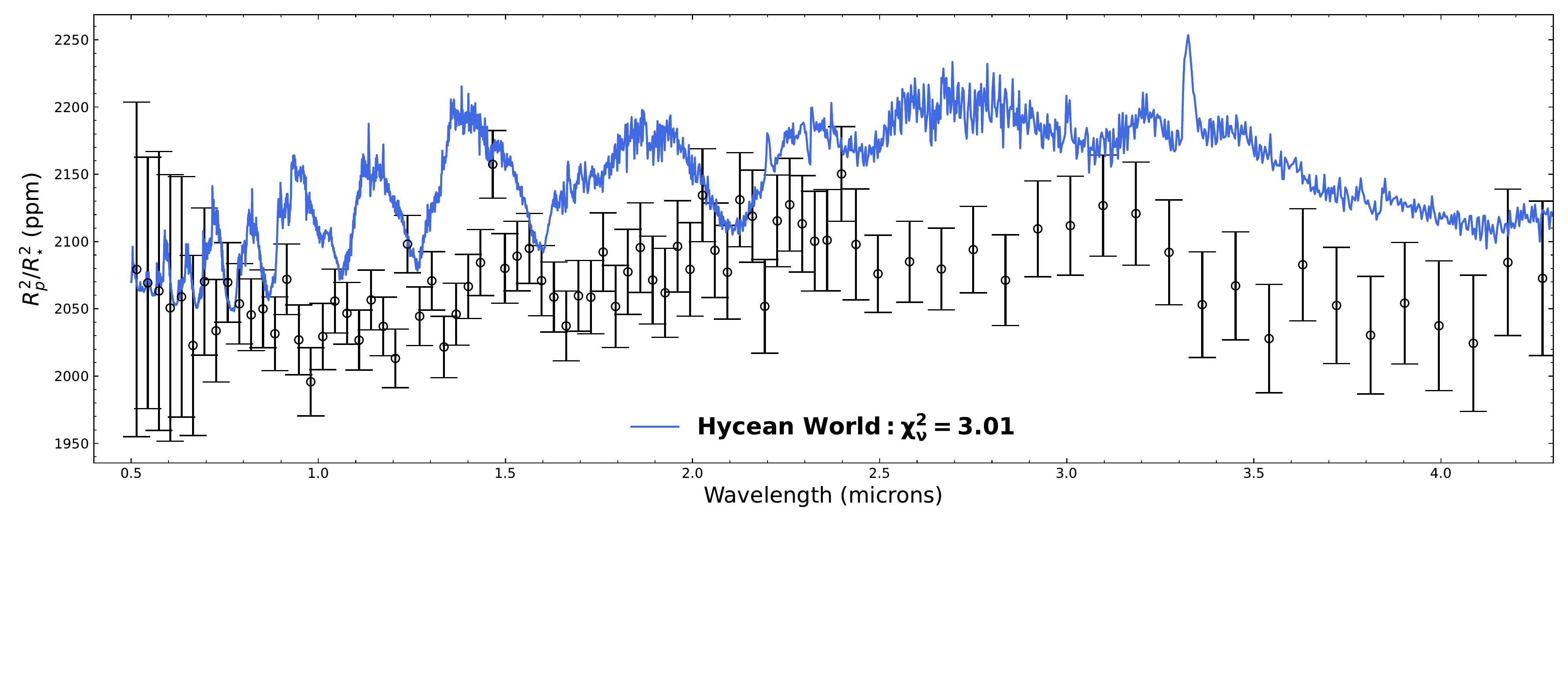}
    \caption{Simulated transmission spectra  of LTT 1445~Ab as a cold Haber world with 90 per cent H$_{2}$ (black points with error bars) compared to a theoretical spectrum of a Hycean world (blue).}
    \label{fig:Chi_squared_comparisionHycean}
\end{figure*}

To quantify if a cold Haber world and Hycean world can be distinguished with \textit{Twinkle}, we implement a $\chi^{2}$ statistical test (Equation \ref{eq:chi}), and sample the \texttt{petitRADTRANS} spectrum  to a common wavelength grid. 
We then calculate the reduced $\chi^{2}$ statistic, $\chi_{\nu}^{2} \equiv \chi^{2}/\nu$, where $\nu$ represents the number of degrees of freedom.  
\begin{equation}
\label{eq:chi}
\chi^{2}=\sum_{i=0}^{n-1}\frac{(Hycean_{i}-Haber_{i})^{2}}{\sigma_{i}^{2}}
\end{equation}
In Equation \ref{eq:chi}, subscript $i$ is the wavelength index to match the Twinkle wavelength grid, Hycean indicates the spectrum for a Hycean World, Haber indicates the spectrum for a cold Haber world, and $\sigma$ is the expected noise for 25 transits observed by \textit{Twinkle}. Based on this metric we find a $\chi_{\nu}^{2} = 3.01$, indicating that \textit{Twinkle} can distinguish between a cold Haber world and a Hycean world (Figure \ref{fig:Chi_squared_comparisionHycean}).
\subsection{Composition of LTT 1445 Ab}

\begin{figure}
    \centering
    \includegraphics[width=\columnwidth]{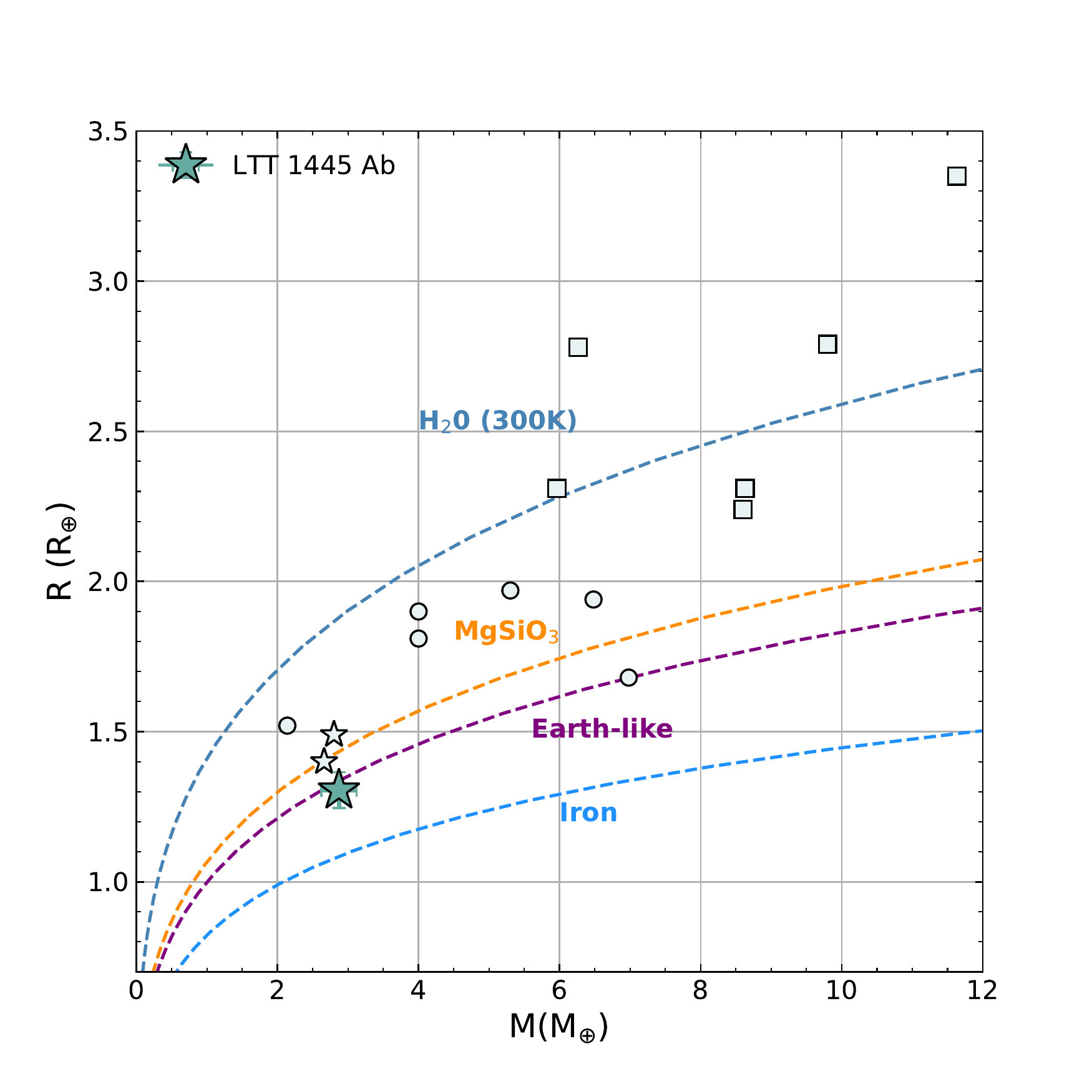}
    \caption{Mass-radius diagram for future targets of interest along with LTT 1445 Ab (blue star). The star markers correspond to objects below the radius valley (R < 1.5 $R_{\oplus}$). The circle markers correspond to objects within the radius valley (1.5 < $R_{\oplus}$ < 2.0). The square markers are those targets above the radius valley ($R_{\oplus}$ > 2.0).  LTT 1445 Ab falls near the composition curve corresponding to an Earth-like planet. The curves are interior structure models of 100 per cent water (dark blue), 100 per cent magnesium silicate rock (orange), 33 per cent iron plus 67 per cent rock (purple) (i.e., Earth-like), and 100 percent iron (light blue). The values are from \citealt{Zeng2019}.}
\label{fig:mass-radius}
\end{figure}

With a planet mass and radius uncertainty of 9 per cent and 5 per cent~\citep{Winters2021}, LTT 1445 Ab is among the best characterized small planets. Such precision in its mass and radius allows us to place constraints on its composition. In Figure~\ref{fig:mass-radius}, we show theoretical mass-radius composition curves~\citep{Zeng2019} and place LTT 1445 Ab in the context of other small exoplanets from the literature that are within \textit{Twinkle's} field of view. LTT 1445 Ab falls near the Earth-like composition curve of 67 per cent magnesium silicate and 33 per cent iron, suggesting that it is likely a rocky planet without a substantial water layer.

We further explore the interior composition of LTT 1445 Ab and calculate its core mass fraction (CMF) using the {\tt ExoPlex} software \citep{Unterborn:2018, Schulze:2020}, which solves the equations of planetary structure and calculates a CMF for a given planet mass and radius. {\tt ExoPlex} assumes a two-layer planet consisting of an iron core and a pure, magnesium silicate ($\rm MgSiO_{3}$) mantle. Assuming the planet mass and radius from \citet{Winters2021} of $R_{p} = 1.304^{+0.067}_{-0.060}~R_{\oplus}$ and $M_{p} = 2.87 \pm 0.25~M_{\oplus}$, we obtain a CMF of $\rm CMF = 0.42^{+0.18}_{-0.17}$. This value is consistent with the value of  $\rm CMF = 0.42 \pm 0.28$ reported by \citet{Winters2021}, calculated using the semi-empirical relations of \citet{Zeng:2017}. As an alternative check on the CMF, we calculated the core radius fraction (CRF) using {\tt HARDCORE} \citep{Suissa2018}, and obtain a value of $\rm CRF = 0.67 \pm 0.14$, which can be easily converted to a CMF using the empirical relations from \citet{Zeng2016}. From that, we obtain $\rm CMF = 0.45 \pm 0.08$, which is consistent with the value from {\tt Exoplex}.

\begin{figure}
    \centering
    \includegraphics[width=1.0\columnwidth]{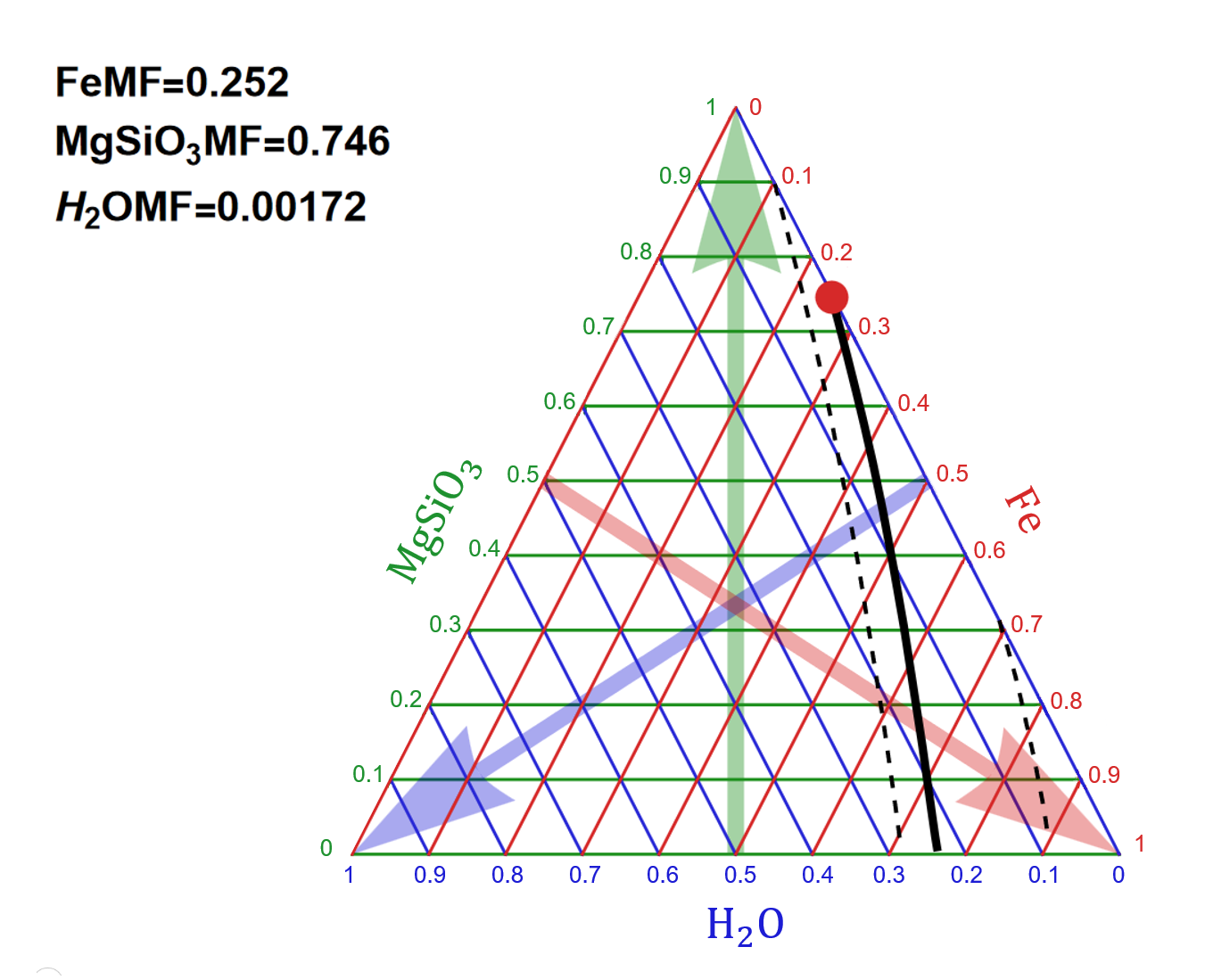}
    \caption{Ternary diagram showing the possible compositions of LTT 1445 Ab using the models from \citet{Zeng:2013}. The red dot denotes relative mass fractions (MF) of the three layers (Fe, $\rm MgSiO_{3}$, $\rm H_{2}O$). The solid black line represents all the possible mass combinations of these layers allowed by the planet's mass and radius, and the black dashed lines represent the uncertainties. The best-fit solution indicates that LTT 1445 Ab has a negligible fraction of water, likely ruling it out as a Hycean world. }
    \label{fig:ternary}
\end{figure}

We also investigated the interior composition of LTT 1445 Ab using the theoretical models of \citet{Zeng:2013}. In these models, the planet consists of three layers: an iron core, a magnesium silicate mantle ($\rm MgSiO_{3}$), and a water layer overlaying the mantle and iron core. Figure~\ref{fig:ternary} shows a ternary diagram with the range of compositions allowed within the uncertainties of the mass and radius of LTT 1445 Ab. We obtain an iron mass fraction of 0.252, a silicate mass fraction of 0.746, and a water mass fraction of 0.002. Some of the other possible solutions along the black line are disfavored for theoretical reasons. For example, a planet consisting purely of water and iron is physically unlikely ($\sim$25 per cent $\rm H_{2}O$ and $\sim$75 per cent Fe core), e.g., \cite{Marcus:2010}. The best-fit solution indicates that LTT 1445~Ab is likely a dry planet, i.e., not a Hycean world as proposed by \citet{MadhusudhanHyean2021}. According to \citealt{MadhusudhanHyean2021}, a typical Hycean planet would have an H$_{2}$-rich atmosphere and a H$_{2}$O layer  with a water mass fraction between 10 per cent and 90 per cent, and iron core + mantle with at least a 10 per cent mass fraction. 
\par
The discrepancy between the classification of \cite{MadhusudhanHyean2021} and ours is probably due to the lower mass they used of $\sim$2.2 $M_{\oplus}$. The revised larger mass with lower uncertainty reported by \citet{Winters2021} leads to a higher density and a more Earth-like, rocky composition, thus ruling out the presence of a thick ocean layer. 
\section{Main Results on NH$_{3}$ Detection}
\label{sec:mainresults}
\subsection{What fraction of hydrogen is Twinkle sensitive to?}
Small planets (R $\lessapprox$  1.6 R$_{\oplus}$) have less gravity and can be prone to losing their atmospheres. Atmospheric loss can be due to either core powered atmospheric mass loss~(\citealt{Gupta_core_mass_2021} and references therein) and/or photoevaporation atmospheric mass loss~(e.g. \citealt{Lopez_Fortney_2013}; \citealt{OwenWu2017}; \citealt{Ginzburg2018}).

\par
\begin{figure}
    \centering
    \includegraphics[width=1.1\columnwidth]{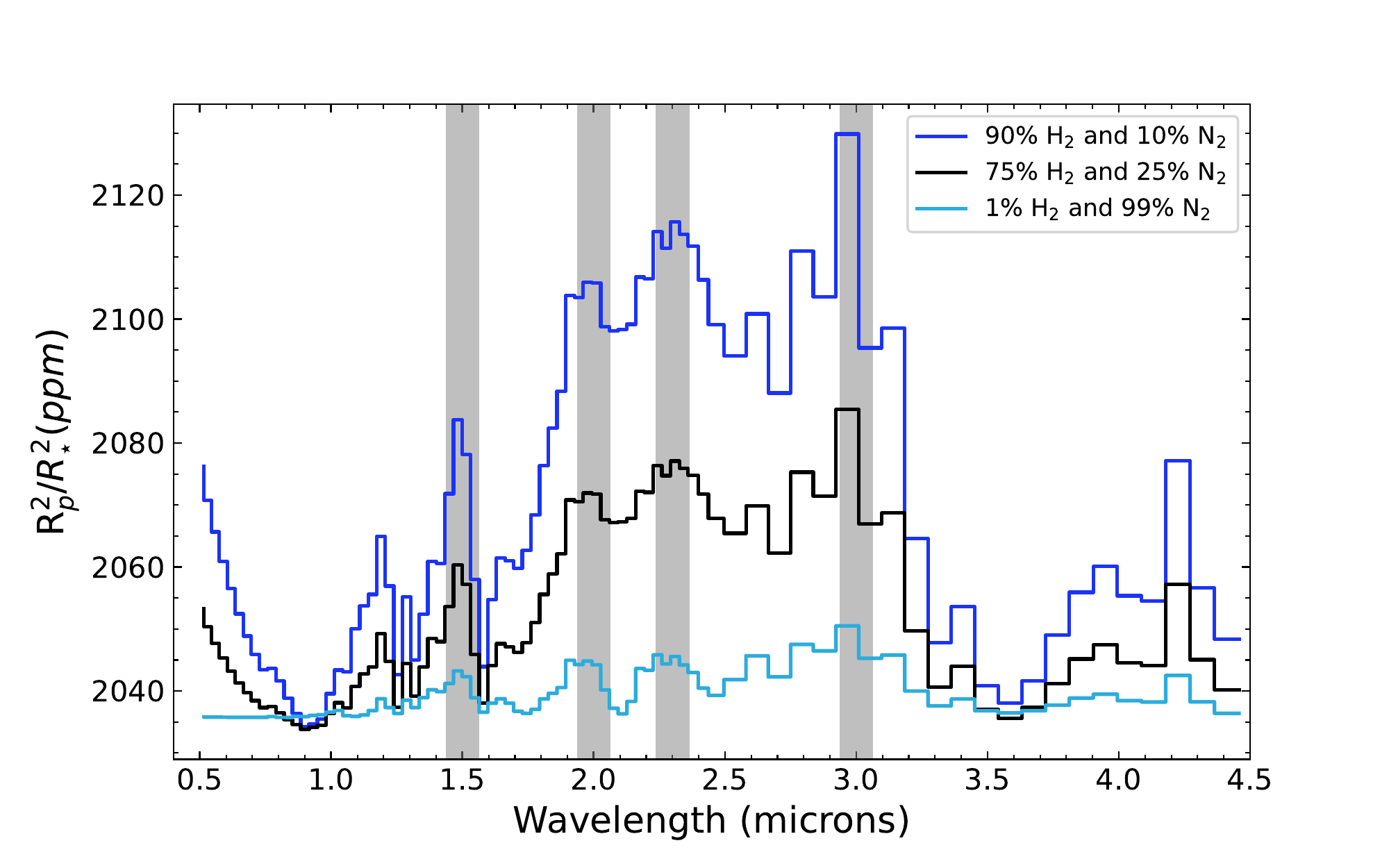}
    \caption{Modeled transmission spectra of LTT 1445 Ab showing various atmospheric compositions compositions. The lines represent different hydrogen dominated scenarios: a H-rich atmosphere (90 per cent H$_{2}$ and 10 per cent N$_{2}$), a H-poor atmosphere (1 per cent H$_{2}$ and 99 per cent N$_{2}$), and a H-intermediate atmosphere (75 per cent H$_{2}$ and 25 per cent N$_{2}$). Contributing NH$_{3}$ features are shown in grey.}
    \label{fig:vary_hydro}
\end{figure}

 
We explore the scenario of  likely H$_{2}$ mass loss of LTT 1445 Ab and see which lower limit fraction of hydrogen \textit{Twinkle} is sensitive to. We follow \citealt{Chouqar2020} and \citealt{Phillips2021} and  consider the following scenarios: a hydrogen-rich atmosphere (90 per cent H$_{2}$ and 10 per cent N$_{2}$), a hydrogen-poor atmosphere (1 per cent H$_{2}$ and 99 per cent N$_{2}$) 
and a hydrogen-intermediate atmosphere (75 per cent H$_{2}$ and 25 per cent N$_{2}$)(Figure \ref{fig:vary_hydro}).   
\par
We determine the effects of a reduction in hydrogen in the atmosphere
on  the detection of ammonia. For LTT 1445~Ab we find that the atmosphere would need to be H-rich (90 per cent H$_{2}$) to be detectable by \textit{Twinkle} (Table \ref{tab:ranking_atm}).

\begin{table}
\centering
\caption{Average S/N of major NH$_{3}$ transmission features for LTT 1445~Ab  transmission spectroscopy for different atmospheric compositions.}
\label{tab:ranking_atm}
\begin{tabular}{cccc}
\hline\hline
LTT 1445 Ab & Ammonia Feature & S/N & Total <{S/N}> 
\\ 
\\
&[$\mu$m]& [$\sigma$]&[$\sigma$]\\
\hline
\multirow{4}{*}{H-rich}&1.5&1.13&\multirow{4}{*}{3.10}\\
&2.0&1.56\\
&2.3&1.75\\
&3.0&1.67\\
\hline
\multirow{4}{*}{H-intermediate}&1.5&0.69&\multirow{4}{*}{1.79}\\
&2.0&0.89\\
&2.3&0.99\\
&3.0&0.97\\
\hline
\multirow{4}{*}{H-poor}&1.5&0.61&\multirow{4}{*}{1.08}\\
&2.0&0.47\\
&2.3&0.49\\
&3.0&0.57\\
\hline
\end{tabular}
\end{table}
\subsection{Other factors that impact NH$_{3}$ detection: ammonia concentration and clouds}

\par 
The lifetime and concentration of  NH$_{3}$ in a H$_{2}$ dominated atmosphere has been previously studied~(e.g. \citealt{Tsai_surface_nh32021}; \citealt{Rajan2022}). We explore how the concentration of ammonia affects the S/N detection in the atmosphere of LTT 1445~Ab.

\citealt{Tsai_surface_nh32021} explored the evolution of the column mixing ratio for NH$_{3}$ for an atmosphere with 1-bar surface with a planet around a quiet M-dwarf host and  a planet around an active M-dwarf host. They found that for a quiet M-dwarf, the mixing ratio of NH$_{3}$ can vary from $\sim$ 10$^{-2}$  to 10$^{-4}$ given a span of 10$^{3}$ to 10$^{8}$ years. In contrast, the atmospheric NH$_{3}$ mixing ratio can vary from $\sim$ 10$^{-2}$ to 10$^{-10}$ around an active M-dwarf over the same span of time. 
\par

In a recent study by \cite{Rajan2022} they found that an Earth-sized planet with an H$_{2}$-dominated atmosphere can enter photochemical runaway of NH$_{3}$ if the net surface production of NH$_{3}$ $\geq$ 2 $\times$ 10$^{10}$ cm$^{-2}$s$^{-1}$. Photochemical runaway occurs of NH$_{3}$ occurs when the production rate of NH$_{3}$ exceeds the finite photochemical destruction rate.
Once in photochemical runaway, the mixing ratio of NH$_{3}$ can increase beyond 10$^{-6}$ with concentrations up to 70 ppmv of NH$_{3}$. 
\par

\begin{figure}
\centering
    \includegraphics[width=1.1\columnwidth]{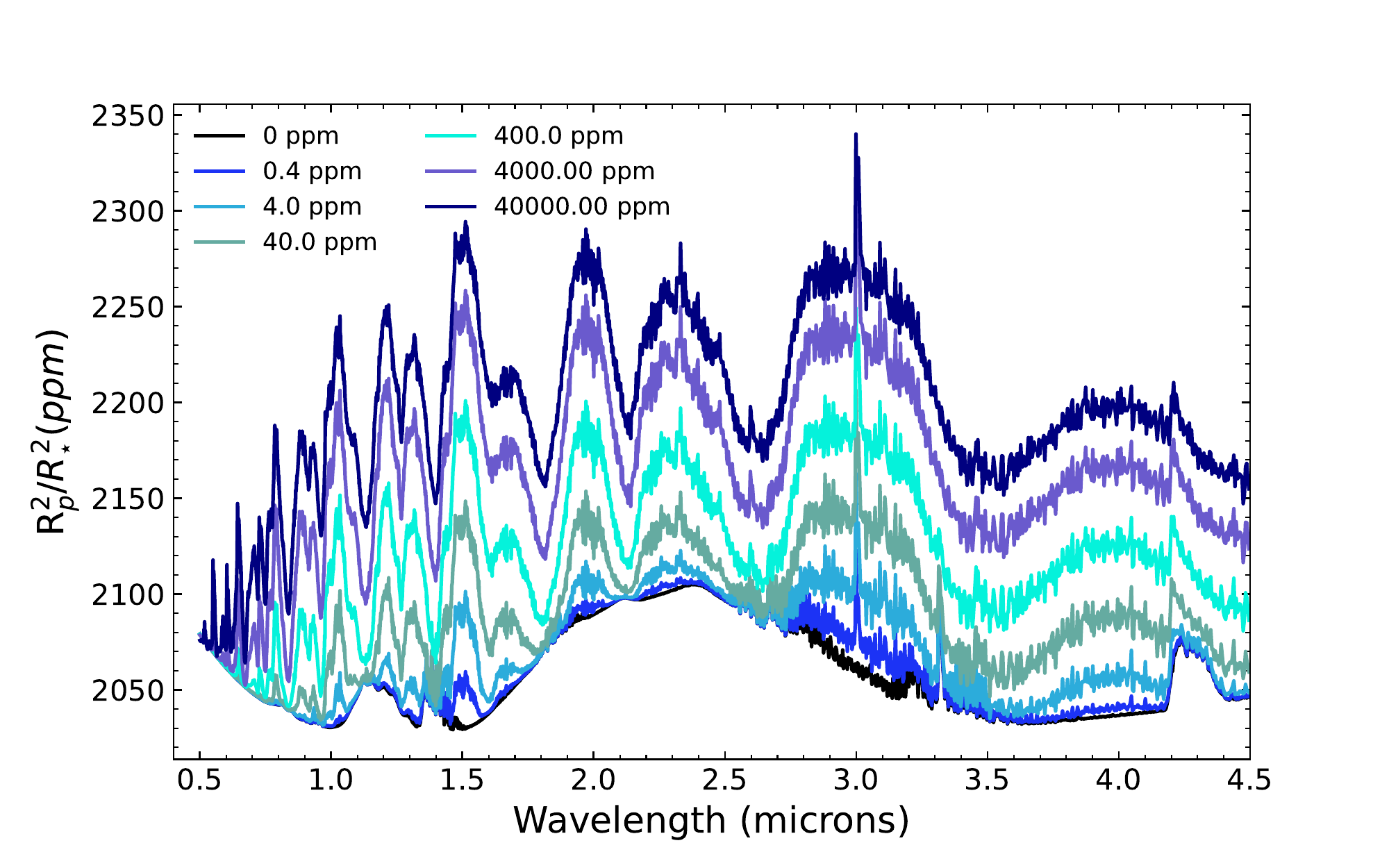}
    \caption{Theoretical transmission spectra of LTT 1445 Ab with varying level of ammonia concentration, 0 ppm, 0.4 ppm, 4.0 ppm, 40 ppm, 400 ppm, 4,000 ppm, and 40,000 ppm}
    \label{fig:vary_ammonia}
\end{figure}

We consider different levels of NH$_{3}$ atmospheric concentration that are within the theoretical range as predicted by \cite{Tsai_surface_nh32021}. We find that a baseline of 4.0 ppm of NH$_{3}$ is needed to be detected by \textit{Twinkle}. Notably, beyond a  concentration of 400 ppm NH$_{3}$, the S/N is nearly constant (Table \ref{tab:varying_ammonia} $\&$ Figure \ref{fig:vary_ammonia}).

\begin{table}
\centering
\caption{NH$_{3}$ transmission features for LTT 1445 Ab transmission spectroscopy for varying concentrations of ammonia}
\label{tab:varying_ammonia}
\begin{tabular}{cccc}
\hline\hline
Concentration of NH$_{3}$ & Ammonia Feature & S/N &  Total <S/N>
\\
\\
&[$\mu$m]& [$\sigma$] & [$\sigma$] \\
\hline
\multirow{4}{*}{0.4 ppm}&1.5&0.10&\multirow{4}{*}{2.55}\\
&2.0&1.52\\
&2.3&1.73\\
&3.0&1.08\\
\hline
\multirow{4}{*}{4.0 ppm}&1.5&1.13&\multirow{4}{*}{3.10}\\
&2.0&1.56\\
&2.3&1.75\\
&3.0&1.67\\
\hline
\multirow{4}{*}{40 ppm}&1.5&2.43&\multirow{4}{*}{4.34}\\
&2.0&1.92\\
&2.3&1.98\\
&3.0&2.30\\
\hline
\multirow{4}{*}{400 ppm}&1.5&3.34&\multirow{4}{*}{5.32}\\
&2.0&2.33\\
&2.3&2.22\\
&3.0&2.59 \\
\hline
\multirow{4}{*}{4000 ppm}&1.5&4.15&\multirow{4}{*}{6.62}\\
&2.0&2.86\\
&2.3&2.66\\
&3.0&3.37 \\
\hline
\multirow{4}{*}{40000 ppm}&1.5&4.30&\multirow{4}{*}{6.75}\\
&2.0&2.94\\
&2.3&2.67\\
&3.0&3.37\\
\end{tabular}
\end{table}
\par 

Additionally, we study the impact of clouds on NH$_{3}$ detection because clouds are known to exist (e.g. \citealt{Kreidberg2014}; \citealt{Helling2019}) and affect transmission spectroscopy observations (e.g. \citealt{Kitzman2010}; \citealt{Benneke2019b}). We use \texttt{petitRADTRANS} to model the effects of clouds by setting a gray cloud deck at 1.0, 0.1, and 0.01 bar (Table \ref{tab:ranking_clouds}). We choose these gray cloud deck levels because condensation curves for temperate exoplanets indicate that H$_{2}$O should condense at pressures below 1.0 bar and form clouds (e.g.~\citealt{Lodders2003}; \citealt{Marley2015temperature}; \citealt{Tinetti2018}). We find that the presence of clouds even at 1.0 bar lowers the S/N of previously observable NH$_{3}$ features to below 3$\sigma$.
\par
\begin{table}
\centering
\caption{Average S/N of major NH$_{3}$ transmission features for LTT 1445~Ab transmission spectroscopy for varying cloud decks with a H-rich atmosphere}
\label{tab:ranking_clouds}
\begin{tabular}{cccc}
\hline\hline

LTT 1445 Ab &Ammonia Feature & S/N & Total <S/N> 
\\ 
\\
&[$\mu$m]&[$\sigma$]&[$\sigma$]\\
\hline

\multirow{4}{*}{Cloud deck at 0.01 bar}&1.5&0.03&\multirow{4}{*}{0.15}\\
&2.0&0.05&\\
&2.3&0.06\\
&3.0&0.12&\\
\hline
\multirow{4}{*}{Cloud deck at 0.1 bar}&1.5&0.24&\multirow{4}{*}{0.91}\\
&2.0&0.36&\\
&2.3&0.52\\
&3.0&0.59&\\
\hline
\multirow{4}{*}{Cloud deck at 1.0 bar}&1.5&0.96&\multirow{4}{*}{2.80}\\
&2.0&1.42&\\
&2.3&1.61\\
&3.0&1.51&\\
\end{tabular}
\end{table}

\section{Atmospheric Retrieval Results}
\label{sec:retrieval_analysis}
In this section we investigate how the abundance of NH$_{3}$  can be constrained using retrieval analysis. 
We use \texttt{petitRADTRANS} \citep{Molliere2020} and \texttt{PyMultiNest} \citep{Buchner2014} to sample the posteriors. \texttt{PyMultiNest} is the Python version of \texttt{Multinest} for nested sampling~\citep{Feroz2009}. In \texttt{PyMultiNest}, we use 2000 live points. {Modeling parameters, priors, and retrieval results can be found in Table \ref{tab:prior}.} 

\subsection{Atmospheric Retrieval Setup}
We use the simulated cold Haber world \textit{Twinkle} data for LTT 1445~Ab as the input (Figure \ref{fig:transmission_Twinkle} \& Table \ref{tab:superearth_low}). To model the simulated data, we use with the following free parameters: surface gravity, planet radius, temperature for the isothermal atmosphere, cloud deck pressure, and mass mixing ratios for different species that are being considered. 
\par
We conduct retrievals for two model setups: (1) a clear atmosphere, (2) an atmosphere with clouds that are parameterized by a grey cloud deck pressure to assess the impact on clouds on the retrieval.
\subsection{Fixing Cloud Deck and Other Minor Species}
In this case, we use the simulated data for the cloud-free low-mean molecular weight case for LTT 1445 Ab (Figure \ref{fig:transmission_Twinkle}). In the retrieval, we assume the cloud deck at $10^{5}$ bar which assumes an essentially cloud-free case for the retrieval. Given the low abundance/low signal of species other than NH$_{3}$, H$_{2}$O and CH$_{4}$, we fix these species as these species would not be readily detectable. Additionally as with the work in \citealt{Phillips2021}, we want to check if NH$_3$ and H$_2$O can be measured given their overlapping features.

\subsubsection{Flat Priors on log (g) and Planetary Radius (R$_{pl}$)}
\begin{figure*}
    \centering
    \includegraphics[width=0.85\textwidth]{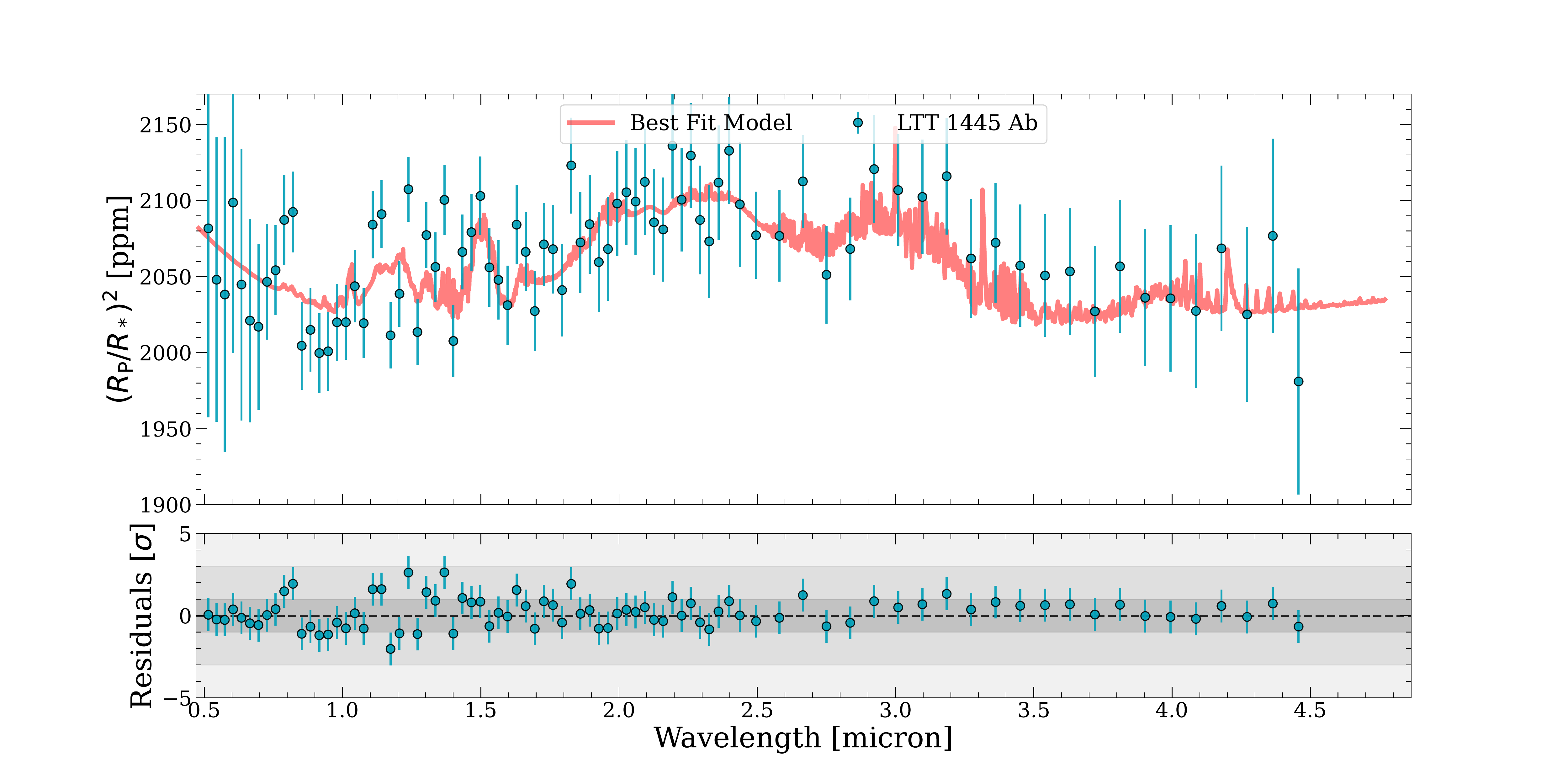}
    \includegraphics[width=0.85\textwidth]{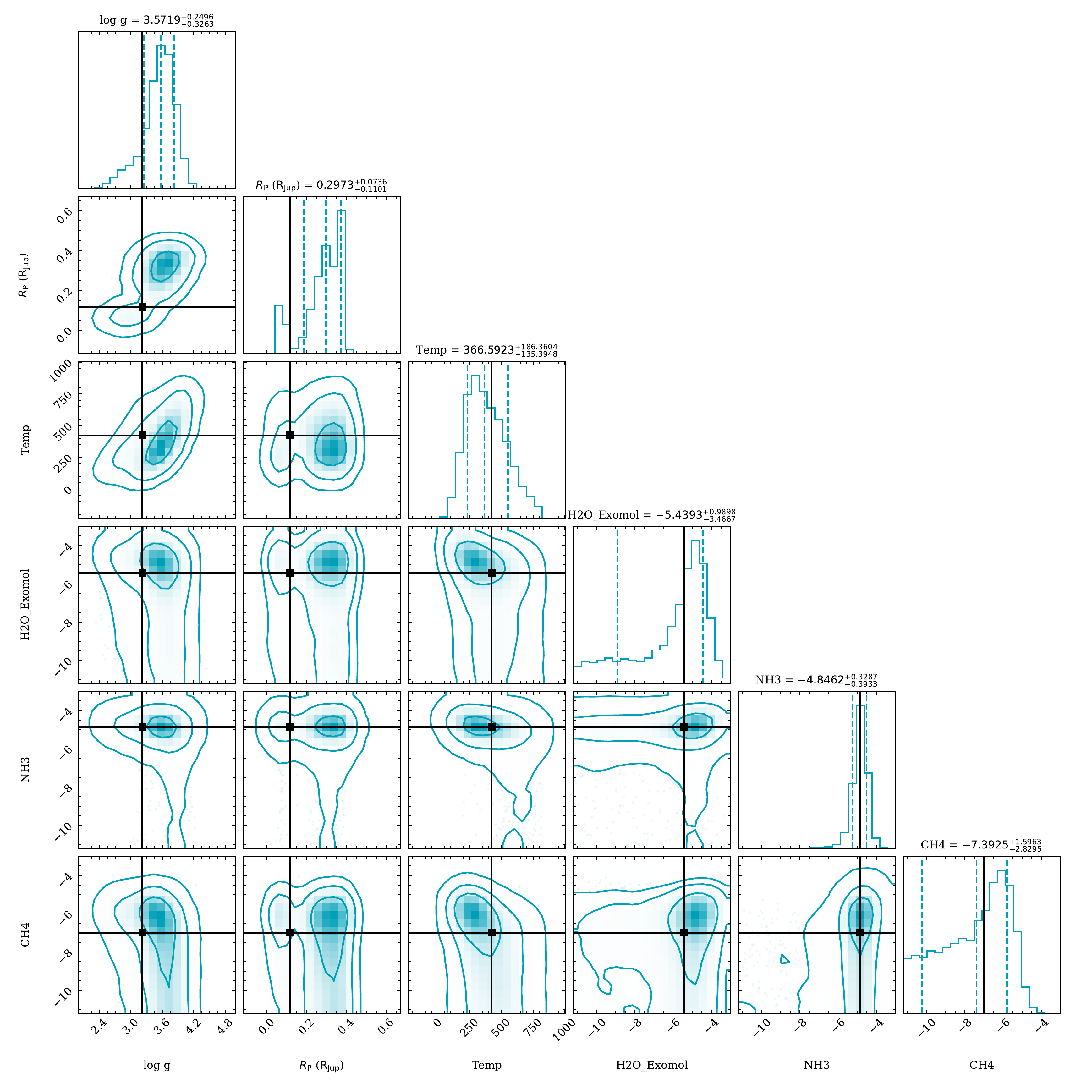}
   \caption{\textbf{Top:} Simulated Twinkle data vs. the retrieved model with flat priors for the surface gravity (log g) and planet radius, and the residuals plotted below. Bottom: Posterior distribution as shown in a corner plot along with the true input values (black lines). Contours are at 1-$\sigma$, 2-$\sigma$, and 3-$\sigma$ from inside out. }
    \label{fig:flat_priors}
\end{figure*}

We apply a flat prior for the surface gravity and planet radius. 
As shown in Figure \ref{fig:flat_priors}, NH$_3$ and H$_2$O can be detected in our retrieval, and their abundances are within 1$\sigma$ from the input values. We note that the planetary radius and log (g) are poorly constrained in the case of flat priors. Given that the radius and mass and thus the surface gravity are more precisely constrained by observations~\citep{Winters2021} we introduce Gaussian priors for these values to test the result of our retrieval analysis.

\subsubsection{Gaussian Priors on log (g) and Planetary Radius (R$_{pl}$)}
\begin{figure*}
    \centering
     \includegraphics[width=0.85\textwidth]{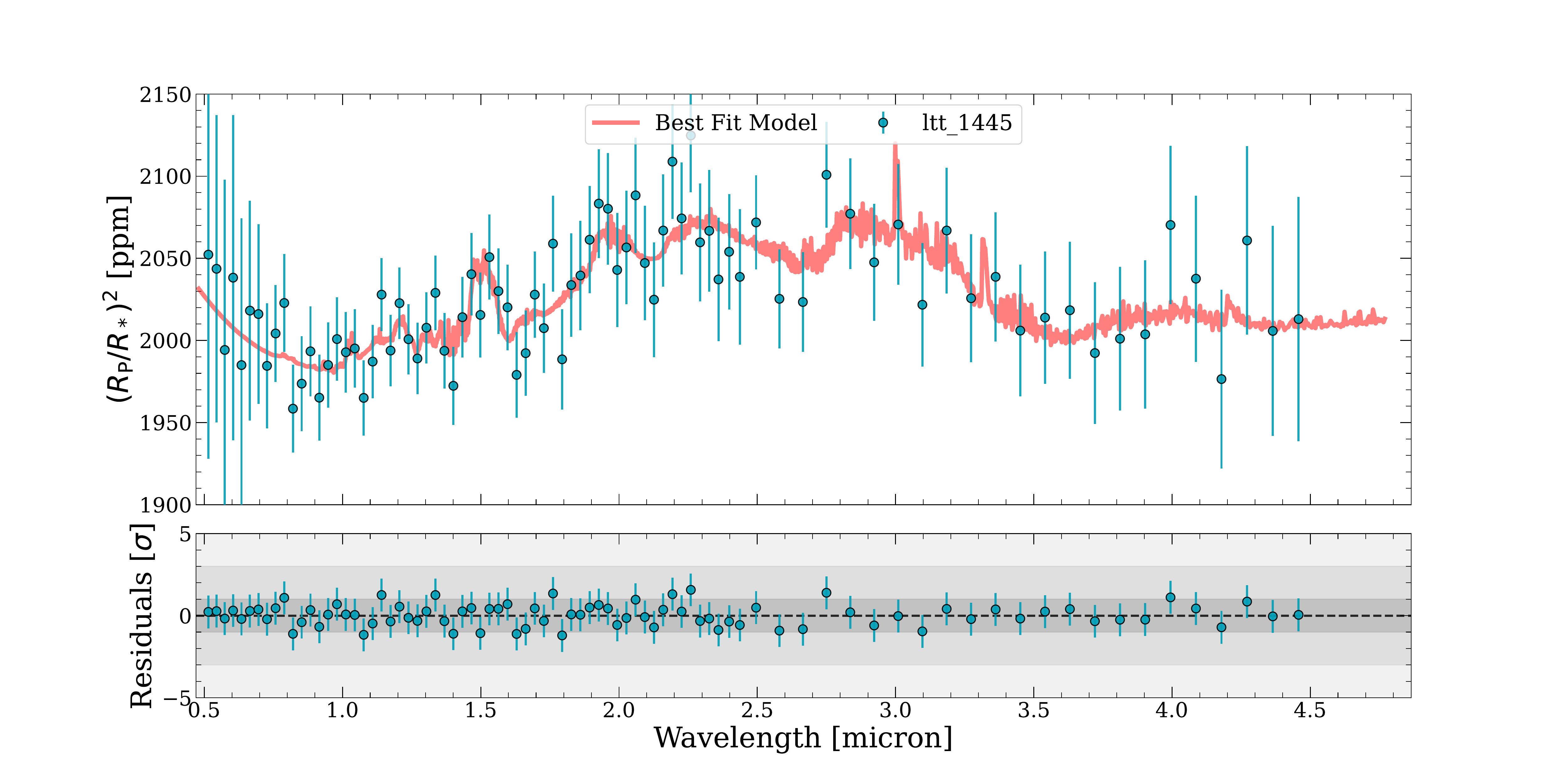}\\
    \includegraphics[width=0.85\textwidth]{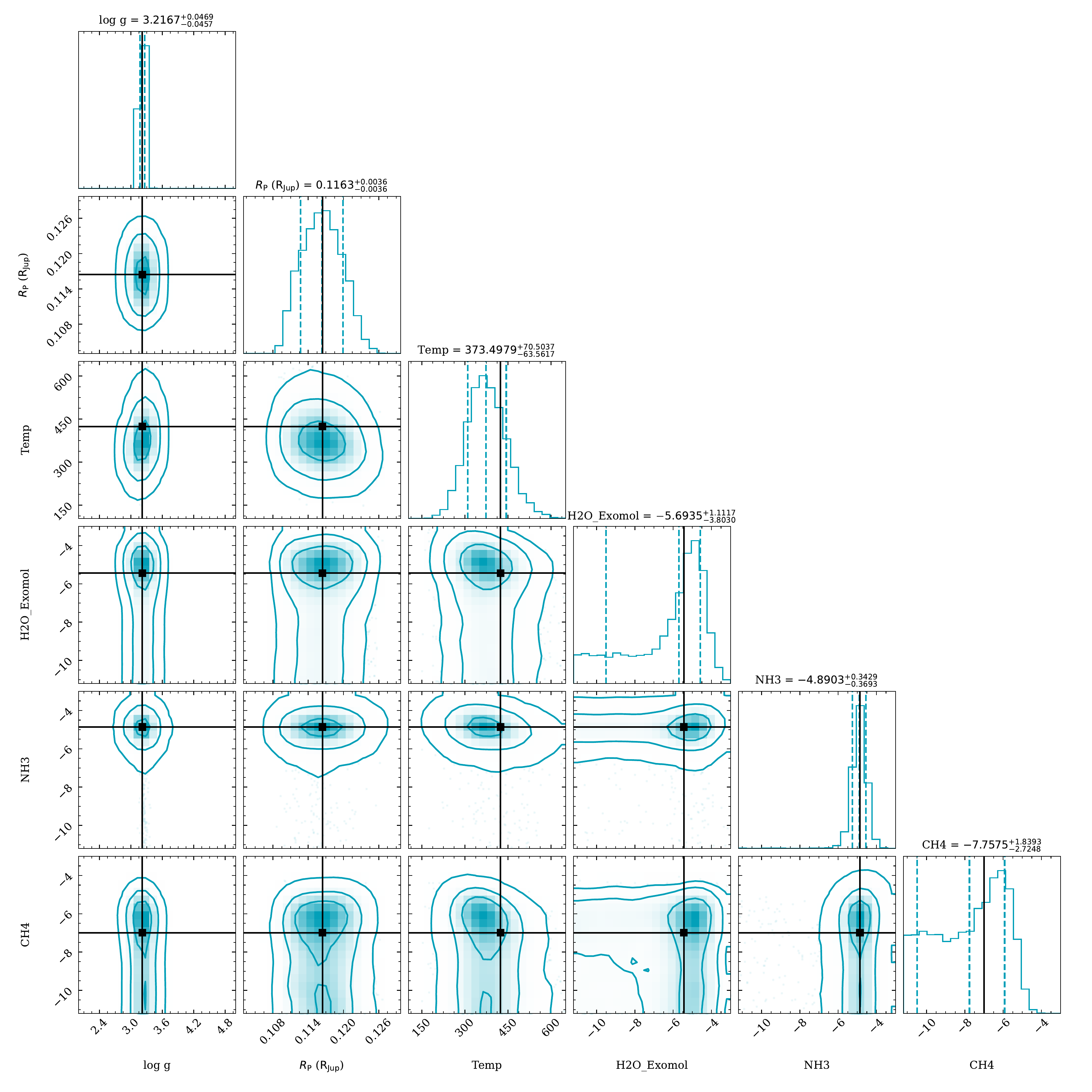}\\
   \caption{Same as Figure. \ref{fig:flat_priors} but with  Gaussian priors for the surface gravity (log g) and planet radius. }
    \label{fig:gaussian_prior}
\end{figure*}

We now consider a retrieval case with Gaussian priors on the log (g) and planetary radius. We apply a Gaussian prior of 3.217$\pm$0.05 dex for log (g) (surface gravity) and 0.1164$\pm$0.005 R$_\mathrm{{Jup}}$ for the planetary radius~\citep{Winters2021}.  In this case, log g and radius are more tightly constrained because of more constraining priors. Additionally, NH$_{3}$ and H$_{2}$O are within 1$\sigma$ of their input values. The corner plot and accompanying spectra are shown in Figure \ref{fig:gaussian_prior}. Retrieval results can be found in Table \ref{tab:prior}. 
\par
To quantify the detection significance, we use similar methods as in \cite{Phillips2021}. Given the 11,611 posterior samples, there are $\sim$0.75 per cent that have a lower value than 10$^{-8}$ mixing ratio for NH$_{3}$. The 10$^{-8}$ mixing ratio threshold is chosen because below this value it is difficult for our retrieval code to constrain abundances (e.g., CH$_4$). The 0.75 per cent fraction translates to 2.6-$\sigma$ assuming a normal distribution. This is consistent with the 3.1-$\sigma$ detection significance from the SNR analysis in \S \ref{sec:mainresults}. 

\
\subsection{Cloud Deck as a Free Parameter}

\begin{figure*}
    \centering
    \includegraphics[width=\linewidth]{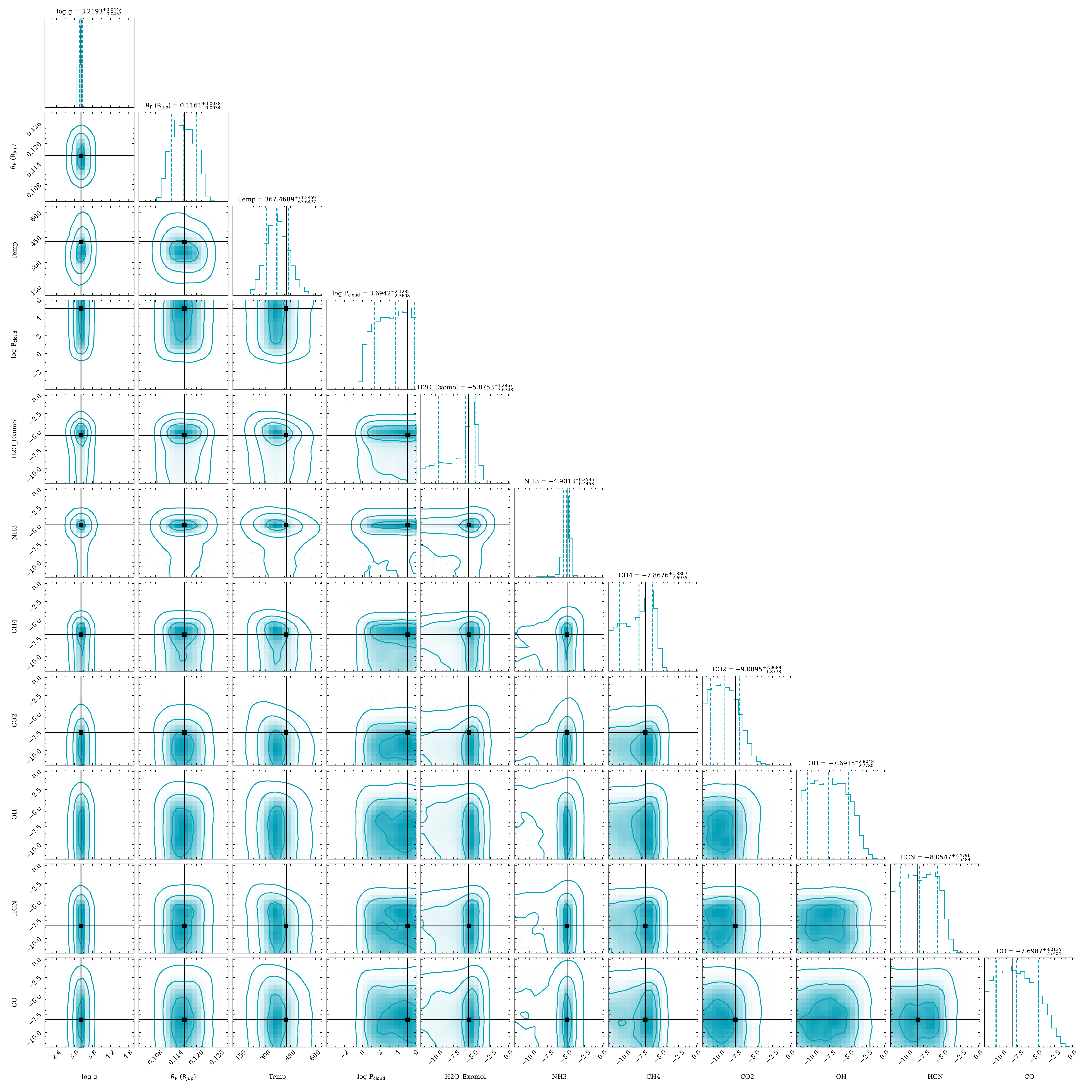} \caption{Corner plot for the full parameter set along with true values (black lines) that are used in generating the \textit{Twinkle} data.}
    \label{fig:full_parameter}
\end{figure*}

Following \cite{Phillips2021}, we also run a retrieval analysis on the full parameter set that includes (1) the cloud deck pressure; and (2) all minor species other than NH$_{3}$, H$_{2}$O and CH$_{4}$. { The prior for the cloud deck pressure covers a range from $10^{-4}$ to $10^{7}$ bar. We are able to put an upper limit to the cloud deck pressure at $\sim$100 bar at at a 1$\sigma$ level. We are unable to constrain minor species with mixing ratio lower than 10$^{-8}$, but we can constrain NH$_{3}$ within 1$\sigma$ of the input value.  The results are in Table \ref{tab:prior} and the corner plot is shown in Figure \ref{fig:full_parameter}}.

\section{Summary and Conclusions}
\label{sec:conclusions}
\par
We model the terrestrial-like planet LTT 1445 Ab for the detection of the potential biosignature ammonia with the upcoming \textit{Twinkle} mission. LTT 1445~Ab is modeled using \texttt{petitRADTRANS} and TwinkleRad. A baseline of 25 transits, 4.0 ppm concentration of NH$_{3}$, and a H-rich atmosphere is considered to determine whether NH$_{3}$ is detectable. 
\par
We demonstrate that \textit{Twinkle} will have the capabilities to distinguish between a cold Haber world and a Hycean world scenario (\S \ref{sec:HabervsHycean}). Given the modeled spectra and the associated uncertainties, we find a $\chi_{\nu}^{2}$ = 3.01, indicating that \textit{Twinkle} can differentiate the two worlds. {Interior composition analysis indicates that LTT 1445~Ab is likely not a Hycean world. This planet is more consistent with a rocky planet without a substantial water mass fraction.}
\par
We explore the fraction of hydrogen needed in the atmosphere of LTT 1445~Ab for ammonia to be detectable (\S \ref{sec:mainresults}). We find that in order to detect NH$_{3}$, LTT 1445~Ab would need a significant portion of H$_{2}$ in the atmosphere (H$_{2}$ = 90 per cent). We also explore the effects on cloud decks and the concentration of NH$_{3}$ on the detectability of NH$_{3}$ in the atmosphere. We find that even the presence of a cloud deck at 1.0 bar would reduce the overall S/N to be lower than 3-$\sigma$ for NH$_{3}$ detection. In addition, we find that a 4.0 ppm concentration of NH$_{3}$ is needed to be detectable by \textit{Twinkle}.

Lastly, we conduct atmospheric retrieval analysis (\S \ref{sec:retrieval_analysis}) which provides helpful insight into constraining NH$_{3}$ and H$_{2}$O  given optimal conditions (i.e a cloud free atmosphere with low MMW). We find that NH$_3$ can be detected at 2.6$-\sigma$. This is consistent with our quadrature sum <S/N> estimate of 3.10$\sigma$.

\par
This work demonstrates that \textit{Twinkle} can provide useful characterization of promising potential smaller terrestrial-like planets to provide insights into potential biosignatures and atmospheric characterization. 

\section*{Acknowledgements}
This research has made use of the NASA Exoplanet Archive, which is operated by the California Institute of Technology, under contract with the National Aeronautics and Space Administration under the Exoplanet Exploration Program.
This project has received funding from the European Union’s Horizon 2020 research and innovation programme under grant agreement No 871149. J.W. acknowledges the support by the National Science Foundation under Grant No. 2143400.
\par
NASA's Astrophysics Data System Bibliographic Services together with the VizieR catalogue access tool and SIMBAD database operated at CDS, Strasbourg, France, were invaluable resources for this work. 
This publication makes use of data products from the Two Micron All Sky Survey, which is a joint project of the University of Massachusetts and the Infrared Processing and Analysis Center/California Institute of Technology, funded by the National Aeronautics and Space Administration and the National Science Foundation.

This work benefited from involvement in ExoExplorers, which is sponsored by the Exoplanets Program Analysis Group (ExoPAG) and NASA’s Exoplanet Exploration Program Office (ExEP).
Caprice Phillips thanks the LSSTC Data Science Fellowship Program, which is funded by LSSTC, NSF Cybertraining Grant $\#$1829740, the Brinson Foundation, and the Moore Foundation; her participation in the program has benefited this work.

This work benefited from the 2022 Exoplanet Summer Program in the Other Worlds Laboratory (OWL) at the University of California, Santa Cruz, a program funded by the Heising-Simons Foundation

This project is supported, in part, by funding from Two Sigma Investments, LP. Any opinions, findings,and conclusions or recommendations expressed in this material are those of the authors and do not necessarily reflects the views of Two Sigma Investments, LP.
This work has made use of data from the European Space Agency (ESA) mission
{\it Gaia} (\url{https://www.cosmos.esa.int/gaia}), processed by the {\it Gaia}
Data Processing and Analysis Consortium (DPAC,
\url{https://www.cosmos.esa.int/web/gaia/dpac/consortium}). Funding for the DPAC
has been provided by national institutions, in particular the institutions
participating in the {\it Gaia} Multilateral Agreement

\begin{landscape}
\begin{table}
\centering
\caption{Parameters used in retrieval, their priors, input and retrieved values.}
\label{tab:prior}
\begin{tabular}{lcccccccc}
\hline\hline
\textbf{Parameter} &
\textbf{Unit} &
\textbf{Type} &
\textbf{Lower or Mean} &
\textbf{Upper or Std} &
\textbf{Input} & 
\multicolumn{3}{c}{\textbf{Retrieved}} \\
 &
 &
 &
 &
 &
 & 
\textbf{Fixed [Gaussian Priors]} & 
\textbf{Fixed [Flat Priors]}&
\textbf{Free [Gaussian Priors]} \\
 \hline
Surface gravity (\textit{logg})&  cgs                 &   Uniform        &  2.0        &   5.0   & 3.217  &\ldots& 3.57$^{     +0.24}_{     -0.33}$ & \ldots \\
Surface gravity (\textit{logg})&  cgs                 &   Gaussian       &  3.217        &   0.050   & 3.217  & 3.21$^{+0.04}_{-0.04}$ &\ldots& 3.21$^{+0.04}_{-0.04}$\\
Planet radius (R$_P$)                       &  R$_{\rm{Jupiter}}$  &   Uniform        &   0.1      &   0.5   & 0.1164 &\ldots&    0.2973$^{  +0.0736}_{  -0.1101}$ & \ldots\\
Planet radius (R$_P$)                       &  R$_{\rm{Jupiter}}$  &   Gaussian        &   0.1164      &   0.005    & 0.1164 &    0.1163$^{  +0.0036}_{  -0.0036}$ &\ldots&    0.1161$^{  +0.00038}_{  -0.00034}$\\
Temperature (T$_{\rm{iso}}$)          &  K                 &   Log-uniform       &  10        &   810     & 424 & 373$^{     +71}_{     -63}$ &366$^{     +186}_{     -135}$& 367$^{     +71}_{     -63}$\\
Cloud pressure ($\log$(P$_{\rm{cloud}}$))           &  bar                 &   Log-uniform       &  -- 4        &   7    & 5.05 & fixed &fixed&       3.69$^{     +2.12}_{     -2.36}$ \\
H$_2$O Mixing Ratio ($\log$(mr$_{\rm{H}_2\rm{O}}$)) &  \ldots            &   Log-uniform      &  --12       &   0   & --5.44  &      --5.69$^{     +1.11}_{     -3.80}$ &--5.43$^{     +0.98}_{     -3.44}$&      --5.87$^{     +1.25}_{     -3.67}$\\
CO Mixing Ratio ($\log$(mr$_{\rm{C}\rm{O}}$))       &  \ldots            &   Log-uniform       &   --12     &   0   & --8.25 & fixed &fixed&      --7.69$^{     +3.01}_{     -2.74}$ \\
CO$_2$ Mixing Ratio ($\log$(mr$_{\rm{C}\rm{O}_2}$)) &  \ldots            &   Log-uniform       &  --12       &   0   & --7.55 & fixed &fixed&      --9.08$^{     +2.06}_{     -1.87}$ \\
CH$_4$ Mixing Ratio ($\log$(mr$_{\rm{C}\rm{H}_4}$)) &  \ldots             &   Log-uniform       &  --12       &   0   & --6.99 &     --7.75$^{     +1.83}_{     -2.72}$ &--7.39$^{     +1.59}_{     -2.82}$&      --7.86$^{     +1.88}_{     -2.69}$ \\
OH Mixing Ratio ($\log$(mr$_{\rm{O}\rm{H}}$)) &  \ldots             &   Log-uniform       &  --12      &   0   & --14.47 & fixed &fixed&      --7.69$^{     +2.80}_{     -2.77}$ \\
NH$_3$ Mixing Ratio ($\log$(mr$_{\rm{N}\rm{H}_3}$)) &  \ldots            &   Log-uniform       &  --12       &   0   & --4.86 &      --4.89$^{     +0.34}_{     -0.36}$ &--4.84$^{     +0.35}_{     -0.39}$&      --4.90$^{     +0.35}_{     -0.44}$ \\
HCN Mixing Ratio ($\log$(mr$_{\rm{H}\rm{C}\rm{N}}$)) &  \ldots            &   Log-uniform       &  --12       &   0   & --8.27 & fixed &fixed&      --8.05$^{     +2.49}_{     -2.54}$ \\
$(R_p/R_{\star})^{2}$ shift ($\Delta_{y}$) &  ppm           &   Uniform       &  --100       &   100   & 0 & 0.0009$\pm$ 0.00012 &-0.0107 $\pm$ 0.0074&      0.0009$\pm$ 0.00012 \\
\hline
\end{tabular}
\end{table}
\end{landscape}

\bibliographystyle{mnras}
\bibliography{main}
\label{lastpage}
\end{document}